\documentclass[conference]{IEEEtran}
\IEEEoverridecommandlockouts

\usepackage{graphicx}
\graphicspath{{./}}
\usepackage{multirow}
\usepackage{amsmath,amssymb,amsfonts}
\usepackage{amsthm}
\usepackage{xcolor}
\usepackage{textcomp}
\usepackage{booktabs}
\usepackage{listings}
\usepackage{array}
\usepackage[numbers,sort&compress]{natbib}
\usepackage{url}

\newcommand{\researchq}[1]{\textbf{RQ#1}}

\begin{document}

\title{Boundary-Mutation Testing for Pattern-Based Secret Detection: A Rule-Level Method and Cross-Scanner Evaluation}

\author{\IEEEauthorblockN{Shweta Mishra}
\IEEEauthorblockA{Independent Researcher\\Moradabad, India\\shweta.mishra.research@gmail.com}}

\maketitle

\begin{abstract}
Pattern-based secret scanners are commonly validated with example-based fixtures that fix one variable: the text surrounding a credential. We introduce boundary-mutation testing to vary that context, generating credentials from each rule's own regular expression, embedding them in realistic source contexts, and classifying outcomes at the rule level rather than the tool level, yielding three detection metrics. Applied to three scanners --- a 43-rule open-source scanner, Gitleaks 8.21.2, and TruffleHog 3.82.13 --- detection in the primary subject holds at $\geq$0.9976 across ten contexts but collapses to 0.5233 when a credential ends in a hyphen. Five rules are affected: two, with fixed-count quantifiers, fail totally and deterministically; three, with variable-count quantifiers, backtrack and match a truncated credential; an entropy fallback rescues some failures but downgrades their severity. We validate a repair restoring full robustness with no new false positives, and report a caution: a plausible first attempt silently regressed two rules, caught only by re-running the same battery. Gitleaks has an unrelated, source-confirmed defect --- a hard-coded terminator allowlist causing total misses for most credential types --- while TruffleHog shows no boundary fragility but narrowest coverage. We report marginal, not conditional, failure probabilities: one common token format is structurally immune, another fails once in 64; on 292{,}527 lines of real code, the false-positive ordering inverts relative to the synthetic corpus. Because per-type detection is deterministic, the comparison unit is ten credential types, not hundreds of samples; no recall difference reaches significance, so we report the null result, not a ranking.
\end{abstract}


\section{Introduction}

Credential leakage through source repositories is a persistent and largely preventable security failure~\citep{meli2019}. The standard control is a pattern-based secret scanner: a table of regular expressions, usually paired with an entropy heuristic, applied to diffs or working trees. Gitleaks~\citep{gitleaks}, TruffleHog~\citep{trufflehog}, GitHub's hosted secret scanning, and the scanner studied here all belong to this family.

Such rule tables are commonly validated with example-based fixtures. A contributor adding a rule writes one or two credentials of the new shape into a test file and asserts they are detected; the scanner we study validates 43 rules with a 313-line fixture suite, and the practice is visible in the test suites of the other two tools as well. We do not claim to have surveyed the field exhaustively, and we do not need to: the weakness we identify follows from the structure of example-based fixtures, not from their prevalence.

That weakness is this. A regular expression does not match a credential in isolation; it matches a credential \emph{together with its surroundings}, and anchoring assertions such as \verb|\b| take effect precisely at the join between the two. Hand-written fixtures almost always place the credential in one canonical position --- inside a double-quoted assignment, ending in an alphanumeric character. A rule that fails only when a credential is passed as a bare function argument, or only when its final character happens to be a hyphen, cannot be observed by a suite in which no fixture is written that way.

\begin{figure*}[t]
\centering
\includegraphics[width=0.92\textwidth]{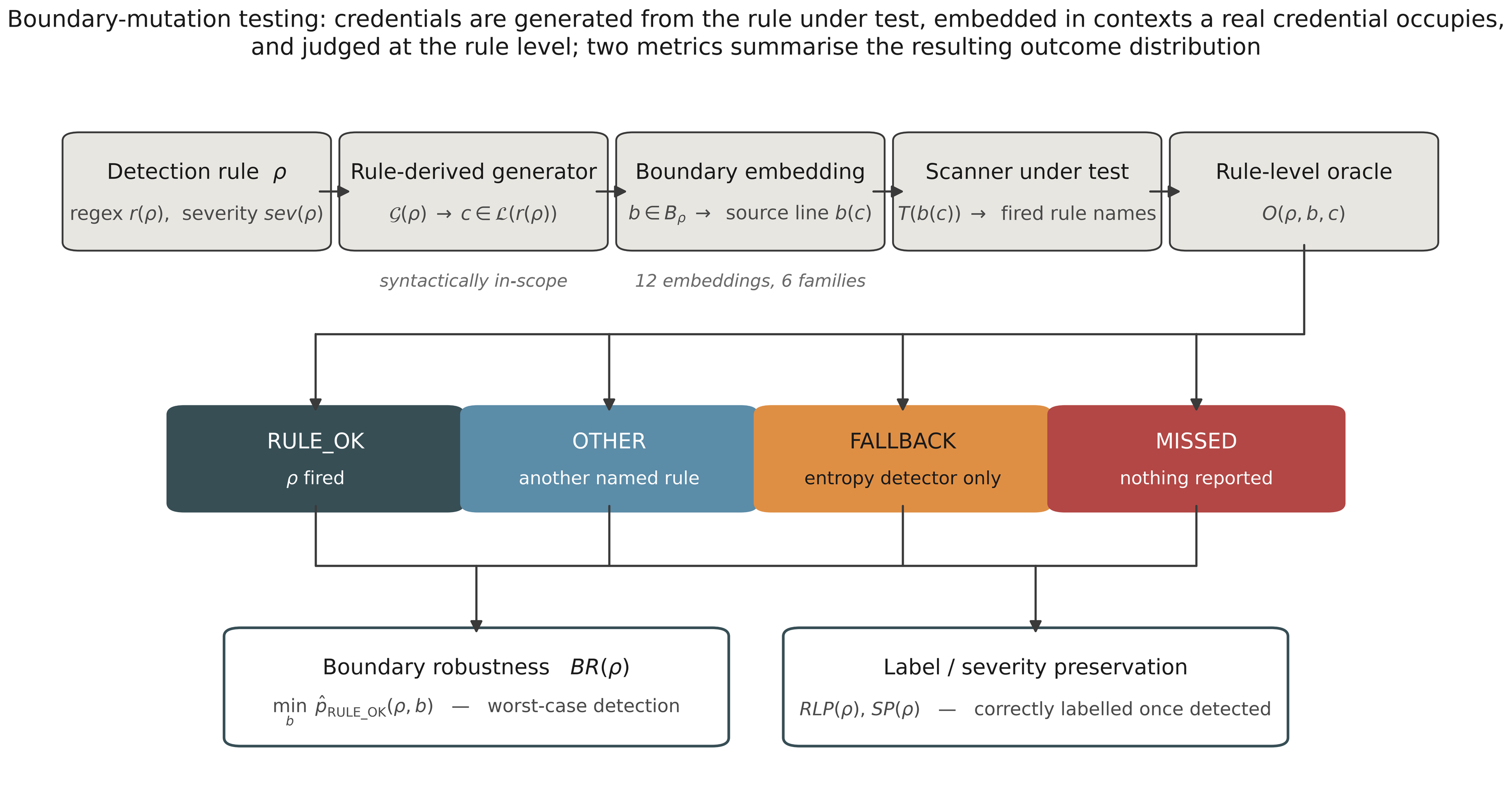}
\caption{Boundary-mutation testing. A credential is generated from the rule under test, embedded in a context that a real credential of that type occupies, and the outcome is classified at the rule level. The four outcome classes yield the three metrics of Section~\ref{sec:metrics}.}
\label{fig:framework}
\end{figure*}

This paper introduces \emph{boundary-mutation testing} to vary that surrounding context systematically. Three design choices distinguish it from writing more test cases. First, credentials are generated from \emph{each rule's own regular expression}, so the corpus is derived from the artefact under test rather than from the experimenter's intuitions. Second, each credential is embedded in a battery of contexts, each one a way a credential legitimately appears in real source. Third --- and this is what makes the defects visible --- outcomes are classified at the \emph{rule} level, distinguishing ``the credential's own rule fired'' from ``something fired'' (Fig.~\ref{fig:framework}). The gap between those two is where a severity downgrade hides, and a tool-level binary metric cannot see it.

We apply the method to three independent scanners to test \emph{transferability and failure-mode diversity}, not to produce a statistically powered ranking; Section~\ref{sec:paired} shows the latter is not supported by ten shared credential types regardless of method. It transfers, and it exposes a distinct scanner-specific behaviour in each: a defect in the first two, and in the third a boundary-robust detector whose limits lie elsewhere. We report the mechanism behind each, the marginal (not merely conditional) probability that each affects a real credential, and a comparative evaluation in which we take some care to avoid disadvantaging a tool whose architecture differs from the others'.

\medskip\noindent\textbf{Contributions.}
\begin{enumerate}
\item \textbf{A method and three metrics.} Boundary-mutation testing, formally defined as a family of metamorphic relations (Section~\ref{sec:method}), with \emph{boundary robustness} $BR(\rho)$, \emph{rule-label preservation} $RLP(\rho)$ and \emph{severity preservation} $SP(\rho)$ derived from its outcome classification.
\item \textbf{Evidence that it generalises.} The same battery applied to three scanners exposes three distinct behavioural profiles (Section~\ref{sec:rq3}), including a previously unreported terminator-allowlist defect in Gitleaks affecting six of the nine credential types it detects.
\item \textbf{Mechanisms, not just findings.} Each defect is traced to a specific regular-expression construct and confirmed on minimal pairs, with marginal failure probabilities computed under an explicitly stated model (Section~\ref{sec:rq2}).
\item \textbf{A validated repair, and a caution about validating repairs.} We patch the five defective rules, re-run the full battery, and confirm $BR=1.0$ with no new false positives on real code. A first candidate patch, plausible on inspection, is shown by the same battery to regress two rules to total failure on a case the original rule handled correctly; we report both results as evidence that a proposed regex repair is a hypothesis, not a fact, until it is run (Section~\ref{sec:repair}).
\item \textbf{Benchmark design as a measured confound.} Corpus construction moves one tool's recall by $+0.10$; deployment condition moves another's detection by $+0.33$; and the real-code false-positive ordering inverts relative to the synthetic corpus (Sections~\ref{sec:rq4}--\ref{sec:rq5}). A paired analysis at the correct unit of observation shows that none of the recall differences we measure is statistically significant, so we report the comparison as a characterisation of trade-offs and not as a ranking.
\end{enumerate}

\medskip\noindent\textbf{Scope.} This is a methods paper about rule-level detection. Two subsystems of the primary subject are therefore deliberately excluded from the main study. Its LLM layer is unmeasured because no funded provider credential was available (Section~\ref{sec:notmeasured}). Its operational characteristics --- queueing, concurrency, and fault tolerance --- were separately characterised and are reported in Online Resource~1; they are outside the scope of the research questions and bear on none of the claims made here.

\medskip\noindent\textbf{Research contribution versus engineering artefact.} Because one of the three evaluated systems is the author's own, we state this distinction explicitly rather than leave it implicit. The \emph{research contribution} is boundary-mutation testing itself: the method, its formalisation as a family of metamorphic relations, its three derived metrics, and the empirical evidence in Sections~\ref{sec:rq1}--\ref{sec:repair} about boundary-sensitive detector behaviour, obtained by applying that method to three independently developed systems. GitHub Autopilot is an \emph{engineering artefact} used to instantiate and validate the methodology --- one experimental subject among three, not the paper's object of study. Nothing in our claims depends on which of the three subjects it is; Section~\ref{sec:bias} details the controls taken because it happens to be ours.

\section{Research Questions}

\noindent\researchq{1} \emph{Does rule-level detection depend on the source context surrounding a credential, and which contexts degrade it?}

\noindent\researchq{2} \emph{What mechanism produces any context sensitivity found, and what is the marginal probability that it affects a real credential?}

\noindent\researchq{3} \emph{Does the method transfer to independently developed scanners, and is the defect class shared or tool-specific?}

\noindent\researchq{4} \emph{How do the three scanners compare on identical corpora, and how much do corpus construction and deployment condition affect that comparison?}

\noindent\researchq{5} \emph{What is the false-positive behaviour of each scanner on the evaluated corpus of unmodified real open-source code?}

\section{Related Work}

\textbf{Secret detection.} \citet{meli2019} measured secret leakage on GitHub at scale, establishing that leaked credentials are common and long-lived. Subsequent work targets the false-positive burden that impedes adoption~\citep{saha2020,rahman2019}, and recent studies examine the accuracy of detection tooling in practice~\citep{basak2022}. To the best of our knowledge, no prior study systematically varies the \emph{context} surrounding a credential while attributing outcomes at the rule level, and we are not aware of prior reports of either defect in Section~\ref{sec:rq3}. Table~\ref{tab:relatedwork} positions this work against the closest related lines.

\textbf{Mutation and metamorphic testing.} Classical mutation testing perturbs the \emph{program} to assess a test suite~\citep{jia2011,demillo1978}. We invert this: the artefact under test is a rule table, and we perturb the \emph{input}. In that respect the method is metamorphic~\citep{chen2018metamorphic,segura2016}: our context mutations are metamorphic relations, since re-quoting a credential or moving it into a YAML file should not change whether it is a credential. The distinguishing elements here are that inputs are generated \emph{from the artefact's own rules}, and that the oracle is rule-level rather than output-level.

\textbf{Input generation and grammar-based testing.} Generating inputs from a formal description relates to grammar-based fuzzing~\citep{godefroid2008} and property-based testing~\citep{claessen2000}, where a generator produces values satisfying a specification. Our generator consumes the detector's own regular expressions, which makes the positive corpus definitionally in-scope for the rule being tested --- a miss cannot be dismissed as outside the rule's syntactic specification, though, as Section~\ref{sec:validity} sets out, syntactic scope is not the same as provider-semantic validity. Work on regular-expression correctness and behaviour~\citep{chapman2016,michael2019} documents that developers frequently misunderstand anchoring and character-class semantics; both defects we report are instances of exactly that, in security-critical rules.

\textbf{Benchmark design as a confound.} Evidence continues to accumulate that benchmark construction, not only technique quality, drives measured outcomes~\citep{liem2020,herbold2020}. Sections~\ref{sec:rq4}--\ref{sec:rq5} contribute three instances in which construction alone reorders tools.

\textbf{Evaluating security scanners.} A recurring finding in the evaluation of static security tools is that measured effectiveness is highly sensitive to the benchmark rather than to the tool: reported detection rates vary widely across corpora, and tools tuned to one benchmark generalise poorly to another. Our Sections~\ref{sec:rq4}--\ref{sec:rq5} are three instances of the same phenomenon in the secret-detection setting, with the additional observation that the sensitivity is large enough to reverse the ordering. Where that literature typically compares tools, we use the sensitivity itself as the object of study and decline to produce a ranking.

\textbf{The unit of analysis in tool comparison.} Our Section~\ref{sec:paired} rests on a point that is well established in empirical software engineering but easy to lose in tool evaluation: when outcomes are deterministic within a stratum, the stratum and not the individual sample is the independent observation. Treating 400 credentials drawn from 10 deterministic types as 400 observations inflates the evidence by the number of samples per type. We report the paired test at the type level for this reason, and consequently report a null result where the sample-level intervals would have suggested a clear ordering.

\begin{table*}[t]
\centering
\caption{This work against the closest related testing methods.}
\label{tab:relatedwork}
\small
\setlength{\tabcolsep}{5pt}
\begin{tabular}{@{}p{5.6cm}ccccc@{}}
\toprule
Line of work & Scan. & Mut. & Meta. & Gen. & Oracle \\
\midrule
Secret-leakage \& FP studies~\citep{meli2019,saha2020,rahman2019,basak2022} & \checkmark & & & & \\
Classical mutation testing~\citep{jia2011,demillo1978} & & \checkmark & & & \\
Metamorphic testing~\citep{chen2018metamorphic,segura2016} & & & \checkmark & & \\
Grammar-/property-based generation~\citep{godefroid2008,claessen2000} & & & & \checkmark$^{*}$ & \\
\textbf{This work} & \checkmark & \checkmark & \checkmark & \checkmark & \checkmark \\
\bottomrule
\end{tabular}
\vspace{2pt}

\raggedright\footnotesize
Scan.~=~evaluates a security scanner; Mut.~=~mutation testing; Meta.~=~metamorphic testing; Gen.~=~inputs generated from a formal rule/specification; Oracle~=~outcome judged at rule level. $^{*}$Generates from a formal specification, not from a detector's own rule. Regex-correctness studies~\citep{chapman2016,michael2019} motivate Sections~\ref{sec:rq2}--\ref{sec:rq3} but are an empirical study, not a method, so are cited in text only and omitted from the rows above.
\end{table*}

\textbf{Standards and formats.} The defects reported here affect detection of JSON Web Tokens~\citep{rfc7519} and OAuth bearer credentials~\citep{rfc6749}, whose encodings determine the marginal probabilities of Section~\ref{sec:rq2}; the HMAC construction underlying JWT signatures is specified in~\citep{rfc2104}. Complexity~\citep{mccabe1976} and coverage~\citep{zhu1997coverage} appear only in the supplementary subject characterisation, and the resilience patterns exercised there are due to \citet{nygard2018release} and \citet{fowler2014circuitbreaker}. \citet{wessel2018bots} and \citet{vasilescu2015ci} study bots and CI at the process level; our subject is one such bot, but it is studied here only as the owner of a rule table.

\section{Subject Systems}

The unit of analysis throughout is a \emph{detection rule}, not a tool. We study three rule tables belonging to independently developed scanners.

The primary subject is the secret scanner of GitHub Autopilot, an open-source self-hosted GitHub App available at \url{https://github.com/Shweta-Mishra-ai/github-autopilot} and studied here at commit \texttt{38b2013}. Its scanner comprises 43 regular-expression rules with per-rule severities, plus one unanchored entropy fallback that fires on high-entropy strings no named rule claims. We make no claim about its deployment scale; it is studied because its rule table is open, small enough to characterise exhaustively, and structurally representative of the regex-plus-entropy family.

\begin{figure*}[t]
\centering
\includegraphics[width=0.92\textwidth]{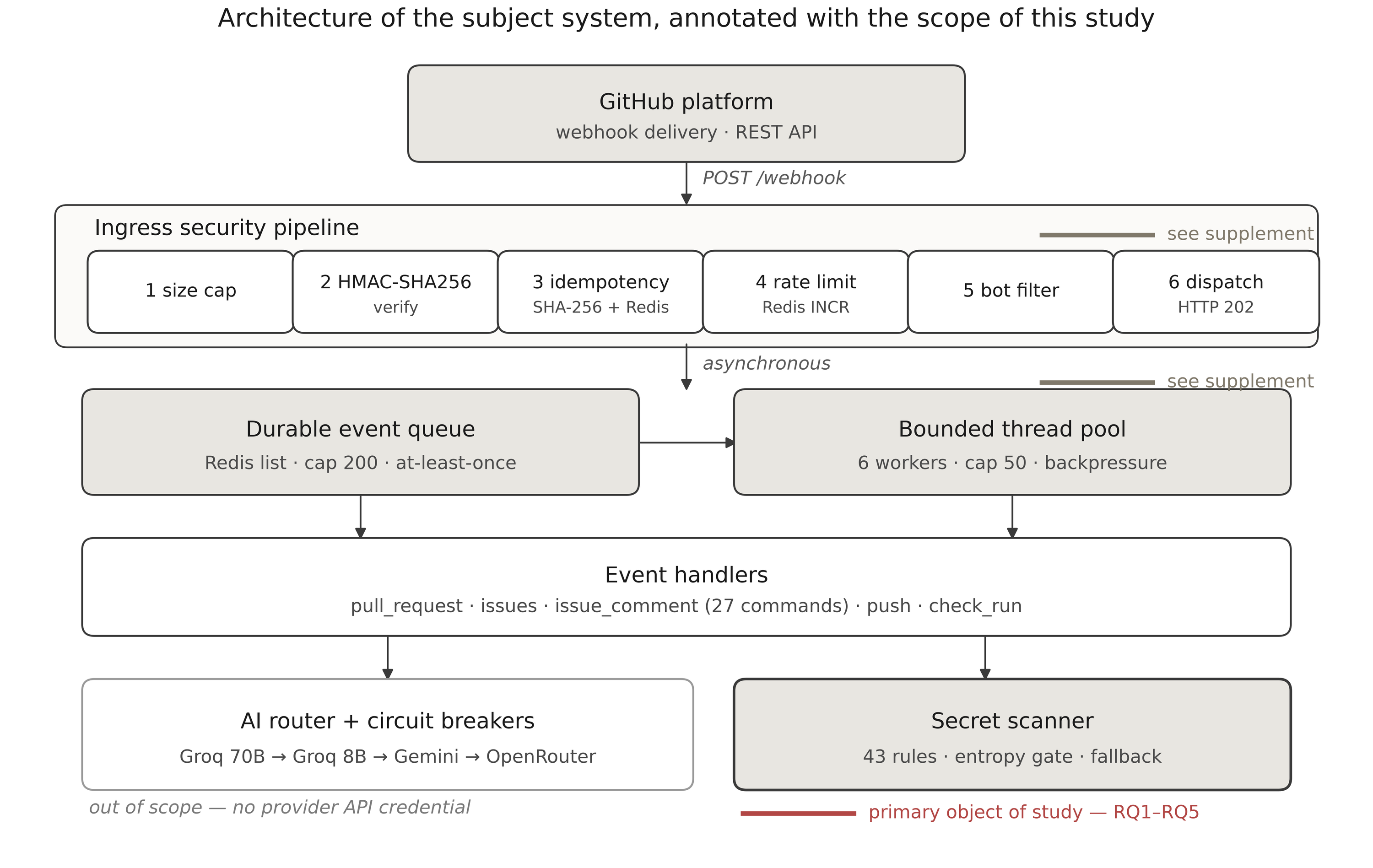}
\caption{The primary subject, annotated with the scope of this study. Only the secret scanner is the object of the research questions; the ingress and concurrency subsystems are characterised in the supplementary artefact, and the LLM routing layer is unmeasured.}
\label{fig:arch}
\end{figure*}

Fig.~\ref{fig:arch} places that scanner within the system it belongs to, and marks the scope of this study. The scanner is the primary object of study. The ingress pipeline, queue and worker pool are characterised in the supplementary artefact but are not the subject of any claim made here. The LLM routing layer is unmeasured entirely (Section~\ref{sec:notmeasured}).

Table~\ref{tab:subject} characterises the subject quantitatively. These measurements describe \emph{what was evaluated}; they are not evidence for or against the correctness of its scanner, and we do not treat them as results. They matter for two reasons only. First, they establish that the subject is a maintained system with a real test suite rather than a toy, which is what makes a defect in its rule table interesting. Second, they establish that its rule table is small enough to characterise exhaustively --- all 43 rules were examined, not a sample.

\begin{table}[t]
\centering
\caption{Characteristics of the primary subject at commit \texttt{38b2013}. Provided to characterise the experimental subject; not treated as evidence of scanner correctness.}
\label{tab:subject}
\small
\begin{tabular}{@{}lr@{}}
\toprule
Property & Value \\
\midrule
Application modules & 91 \\
Application lines of Python & 19{,}352 \\
Test lines of Python & 19{,}825 \\
Functions and methods & 617 \\
Mean cyclomatic complexity & 4.67 \\
Maximum cyclomatic complexity & 34 \\
\midrule
Test cases & 2{,}054 \\
Test outcome & 2{,}054 passing \\
Statement coverage & 83.9\% (7{,}600 statements) \\
Lint violations (\texttt{ruff}) & 0 \\
\midrule
Scanner rules & 43 named $+$ 1 entropy fallback \\
Rules analysed in this study & 24 (Section~\ref{sec:method}) \\
Scanner fixture suite & 313 lines \\
\bottomrule
\end{tabular}
\end{table}

The last two rows are the ones that motivate this paper. A 43-rule table validated by a 313-line fixture suite is not unusual, and it is not negligent; it is the normal ratio in this class of tool. Section~\ref{sec:rq1} shows what that ratio leaves unobserved.

The comparison subjects are Gitleaks 8.21.2 and TruffleHog 3.82.13, both widely used in practice and both independently developed by parties unconnected to the primary subject. Gitleaks is rule-table-driven with a public TOML configuration, which makes source-level mechanism confirmation possible; TruffleHog is detector-driven with co-occurrence and live-verification logic, an architecture different enough from the other two that Section~\ref{sec:rq4} treats it as a distinct evaluation condition rather than a like-for-like competitor.

\section{Method: Boundary-Mutation Testing}
\label{sec:method}

This section defines each element of the method summarised in Fig.~\ref{fig:framework}.

\subsection{Definition}

Let $\rho$ be a detection rule with regular expression $r(\rho)$ and declared severity $sev(\rho)$, and let $\mathcal{G}(\rho)$ be a generator producing strings in $\mathcal{L}(r(\rho))$, the language of that expression. Let $B$ be a set of \emph{boundary embeddings}, each $b \in B$ a function mapping a credential $c$ to a line of source text $b(c)$ in which $c$ appears as a real credential would. Let $T$ be the tool under test, returning a set of fired rule names $T(b(c))$.

\subsubsection*{Syntactic versus semantic validity}
\label{sec:validity}

Two notions of ``valid credential'' must be kept apart, because the method guarantees one and not the other. Write $\mathcal{L}(r(\rho))$ for the set of strings the rule's own expression accepts, and $C_\rho$ for the set of strings the issuing provider would actually mint as a credential of that type. In general $C_\rho \subseteq \mathcal{L}(r(\rho))$, and the inclusion is strict: a rule that accepts \texttt{SG.} followed by 22 and then 43 characters from a permitted class also accepts strings that SendGrid would never issue.

Our generator samples from $\mathcal{L}(r(\rho))$, so it guarantees \emph{syntactic} in-scope\-ness only. This is the precise sense in which a miss is a defect: the rule's author declared, by writing $r(\rho)$, that strings in $\mathcal{L}(r(\rho))$ occurring in these contexts are to be reported, and the tool did not report one. That is an internal inconsistency in the rule, and it is what the method detects. It is \emph{not} a claim that the specific string would ever be issued. Where $C_\rho$ is much smaller than $\mathcal{L}(r(\rho))$, a defect found on $\mathcal{L}(r(\rho)) \setminus C_\rho$ may have no field consequence, which is exactly why Section~\ref{sec:rq2} computes marginal probabilities under an explicit issuance model rather than resting on the generated rates alone. Where a provider's format is public and the rule encodes it faithfully --- the case for the five defective rules we report --- the two sets nearly coincide and the distinction is immaterial.

\subsubsection*{The metamorphic relation}
\label{sec:mr}

The mutations are metamorphic relations over the tool's input. For a credential $c$ and two embeddings $b_i, b_j$ both applicable to $\rho$:
\[
\mathit{MR}(c, b_i, b_j):\quad
T\bigl(b_i(c)\bigr) \ni \rho \;\Longleftrightarrow\; T\bigl(b_j(c)\bigr) \ni \rho .
\]
The invariant asserted is not textual equality between $b_i(c)$ and $b_j(c)$ --- they differ by construction --- but \emph{preservation of credential detectability under a semantics-preserving change of surrounding context}. Re-quoting a credential, moving it into a YAML file, or passing it as a call argument does not change whether the string is a credential, so a rule whose verdict changes has violated the relation. A violated relation localises the defect without requiring an independent oracle for what the tool ``should'' output, which is the standard advantage of metamorphic testing and the reason it applies here: there is no ground-truth detector to compare against.

For a credential $c \sim \mathcal{G}(\rho)$ and embedding $b$, the outcome is classified as
\[
O(\rho,b,c)=
\begin{cases}
\textsc{rule\_ok} & \rho \in T(b(c))\\
\textsc{other} & T(b(c)) \setminus (F \cup \{\rho\}) \neq \emptyset\\
\textsc{fallback} & \emptyset \neq T(b(c)) \subseteq F\\
\textsc{missed} & T(b(c)) = \emptyset
\end{cases}
\]
where $F$ is the tool's set of unanchored fallback detectors. An embedding $b$ is \emph{applicable} to $\rho$ when $b$ preserves membership in $\mathcal{L}(r(\rho))$; forcing a final character the rule's own character class excludes would produce a string that is no longer a credential of that type, and a miss there would be correct behaviour rather than a defect. Let $B_\rho \subseteq B$ be the applicable embeddings.

\subsection{Metrics}
\label{sec:metrics}

Writing $\hat p_{\textsc{x}}(\rho,b)$ for the empirical rate of outcome \textsc{x} over $n$ samples, we define three metrics. The first summarises how often the rule works; the other two summarise, of the credentials that were detected at all, how faithfully they were reported.

\emph{Boundary robustness} is the rule's worst-case behaviour over contexts in which it is supposed to work:
\begin{equation}
BR(\rho) = \min_{b \in B_\rho}\ \hat p_{\textsc{rule\_ok}}(\rho,b). \label{eq:br}
\end{equation}

\emph{Rule-label preservation} is the fraction of detected credentials reported under the credential's own rule:
\begin{equation}
RLP(\rho) = \frac{\sum_{b \in B_\rho} \hat p_{\textsc{rule\_ok}}(\rho,b)}
                 {\sum_{b \in B_\rho} \bigl(1-\hat p_{\textsc{missed}}(\rho,b)\bigr)}. \label{eq:rlp}
\end{equation}

$RLP$ measures rule \emph{identity}, which is not the same thing as severity. A tool that reports a credential under a different rule carrying the same severity has mislabelled it without downgrading it, and a metric named for severity should not penalise that. We therefore define severity preservation separately, over the severity actually reported. Let $\widehat{sev}$ denote the severity attached to the finding a tool emits: $sev(\rho)$ for \textsc{rule\_ok}, $sev(\rho')$ for an \textsc{other} outcome firing $\rho'$, and $sev(F)$ for a \textsc{fallback} outcome. Then
\begin{equation}
SP(\rho) = \frac{\sum_{b \in B_\rho} \mathbb{E}\bigl[\mathbb{1}\{\widehat{sev} = sev(\rho)\}\bigr]}
                {\sum_{b \in B_\rho} \bigl(1-\hat p_{\textsc{missed}}(\rho,b)\bigr)}. \label{eq:sp}
\end{equation}

$SP < 1$ means the tool is quietly reporting findings it did detect at the wrong severity; $RLP < 1$ means it is attributing them to the wrong rule. The two coincide only when no substitute label shares the true rule's severity. A rule is \emph{boundary-fragile} when some applicable embedding degrades it materially relative to the canonical one.

Both metrics require a severity oracle. For the primary subject we have one: each rule declares a severity, and the fallback detector's severity is fixed at \textsc{medium} in source. Section~\ref{sec:rq1} reports both. For the two external scanners we do not attempt a severity mapping --- their severity models are not commensurable with the subject's, and inventing a correspondence would manufacture a comparison the data does not support --- so Section~\ref{sec:rq3} reports $BR$ and $RLP$ only.

\subsection{Instantiation}

$\mathcal{G}$ is implemented with \texttt{exrex}, which enumerates strings from a regular expression. Of the 43 rules, seven are structural (PEM headers, connection-string schemes) and twelve are keyword-anchored --- they require an adjacent keyword such as \texttt{aws\_secret}, so a neutral embedding would make them unmatchable by construction --- leaving 24 rules. We use $n=120$ samples per cell.

\subsubsection*{Constructing the boundary battery}

$B$ comprises twelve embeddings drawn from six families (Table~\ref{tab:taxonomy}). The families are not arbitrary: each corresponds to a distinct syntactic mechanism by which a character can come to sit immediately adjacent to a credential in source code, which is the only thing an anchoring assertion can observe. Within a family the members vary the adjacent character while holding the mechanism fixed.

We make no claim that twelve embeddings exhaust the boundary space, and Section~\ref{sec:threats} names the family we know is missing. The battery is a \emph{representative} boundary battery, chosen to cover each mechanism at least once, not an exhaustive enumeration. Its adequacy is argued from mechanism coverage, and is supported by the distinct scanner-specific behaviours it exposed across the three tools to which it was applied.

\begin{table}[t]
\centering
\caption{The boundary battery: six families, twelve embeddings}
\label{tab:taxonomy}
\small
\setlength{\tabcolsep}{3.5pt}
\begin{tabular}{@{}llp{3.25cm}@{}}
\toprule
Family & Embedding & Mechanism exercised \\
\midrule
\multirow{3}{*}{Assignment}
 & quoted & canonical fixture shape \\
 & single-quoted & alternate quote character \\
 & unquoted & no delimiter at all \\
\midrule
\multirow{3}{*}{Structured data}
 & YAML value & configuration syntax \\
 & JSON value & configuration syntax \\
 & dotenv & environment-file syntax \\
\midrule
Expression
 & call argument & credential passed programmatically \\
\midrule
Collection
 & list member & separator-terminated \\
\midrule
URL
 & query parameter & the shape credentials most often leak in \\
\midrule
Lexical
 & extra whitespace & whitespace as terminator \\
\midrule
\multirow{2}{*}{Final character}
 & ends \texttt{\_} & permitted terminal, word character \\
 & ends \texttt{-} & permitted terminal, non-word character \\
\bottomrule
\end{tabular}
\end{table}

\subsubsection*{Freezing the expected rule for external scanners}
\label{sec:freeze}

Judging one tool by another project's rule names would be unfair, so for Gitleaks and TruffleHog the expected rule per credential type is established empirically and frozen before any mutation is applied. For each credential type we (i) generate samples from the primary subject's rule for that type; (ii) present them in the \emph{canonical} embedding only; (iii) record which \texttt{RuleID} or \texttt{DetectorName} the tool fires; (iv) require that the same identifier fire on every one of the canonical samples, so that the expectation is not set by a single observation; and (v) freeze that identifier as the expected rule for all subsequent embeddings. A type on which a tool fires no rule, or fires inconsistently, in the canonical context is excluded from that tool's aggregate and reported separately as a coverage gap, so that missing coverage is never scored as fragility.

The residual risk in this procedure is that the canonical context itself provokes a wrong-but-consistent attribution, which would then be frozen as the expectation. We checked the frozen identifiers by inspection and they are semantically correct in every case (for instance \texttt{aws-access-token} for AWS keys); no type was attributed to a generic or unrelated detector. The procedure cannot detect an error that is both consistent and semantically plausible, which we note as a limitation rather than claim to have eliminated.

\subsection{A control that proved necessary}
\label{sec:control}

An initial harness used the identifier \texttt{SECRET} in its embeddings and measured a uniform 100\% detection rate, apparently exonerating every rule. The identifier was itself triggering the scanner's \emph{generic} keyword rules, which fired on the surrounding text and masked the failure of the specific rule under test. All results below use deliberately neutral identifiers (\texttt{cfg\_value}). We report this because it is the failure mode a replication is most likely to reproduce, and because it is a concrete instance of why the rule-level oracle matters: a tool-level oracle could not have distinguished the two situations at all.

\subsection{What we did not measure}
\label{sec:notmeasured}

The primary subject ships an AI-output evaluation harness requiring a funded provider key; our environment had neither the key nor outbound access to the four providers. We therefore do not measure LLM latency, output quality, or hallucination rate, and substitute no estimates. Nothing in this paper speaks to the subject's AI behaviour.

We also do not treat the subject's operational behaviour as a result of this study. Its queueing, concurrency and fault-tolerance characteristics were measured --- including a lost-update oracle, a synchronised-start scaling curve, a payload sweep and a fault-injection battery --- and are reported in Online Resource~1, with the full raw data in the replication package. They appear as a separate online resource rather than as a research question because they concern a different artefact (a running service) than the one this paper is about (a rule table), and because no claim in this paper depends on them.

\subsection{Researcher bias and evaluation controls}
\label{sec:bias}

The author wrote the primary subject. That is a competing interest with a specific and predictable direction: it would bias the study toward flattering the subject's scanner. We state the controls used against it, and the evidence that they bit.

Procedurally: the subject was frozen at commit \texttt{38b2013} before any benchmark was run and never modified in response to a result; the external tools are unmodified official release binaries at pinned versions, verified by SHA-256; all three tools receive byte-identical corpora and identical execution conditions; every raw output is retained in the replication package, including the runs that contradict the subject; no finding was excluded after inspection; and no defect found in the subject was fixed before measurement.

Substantively, three of this paper's results are unfavourable to the subject and are reported at full strength. The paper's single largest finding is a defect in the subject's own scanner, not in a competitor's. The subject's rules are characterised as \emph{shallower} than Gitleaks' --- prefix, class and length only --- and that shallowness is identified as the cause both of its perfect recall and of its 17 false positives (Section~\ref{sec:rq4}). Every one of the subject's ten real-world findings is a false positive, itemised in Section~\ref{sec:rq5}. Most consequentially, the paired analysis in Section~\ref{sec:rq4} finds that \emph{none} of the recall differences favouring the subject is statistically significant, and we therefore withdraw any ranking claim rather than report the raw ordering as a result.

The design that most constrains this bias is the one built into the method: credentials are generated from \emph{each tool's own} declared rules, so the corpus cannot be tuned to favour a particular tool's shapes without that tuning being visible in the rule table itself.

\section{Experimental Setup}
\label{sec:setup}

\subsection{Environment}

All measurements were produced in a single containerised Linux environment running CPython 3.11.15. The comparison scanners are official release binaries, Gitleaks 8.21.2 and TruffleHog 3.82.13, pinned by version and verified by SHA-256 digest at reproduction time; the digests are recorded in the replication package. TruffleHog is invoked with \texttt{--no-verification}, since live credential verification requires outbound network access to the issuing providers and was not available. Where a Redis instance is required it is \texttt{redis-server} 7.0.15 bound to loopback with persistence disabled.

Hardware is deliberately not reported as a study parameter. Every result in Sections~\ref{sec:rq1}--\ref{sec:rq5} is a detection \emph{outcome} --- which rule fired on which input --- and is therefore invariant to the machine that produced it. The only hardware-sensitive measurements in this work are the operational timings, which are reported in the supplementary artefact and are not the basis of any claim here.

\subsection{Corpora and sampling}

Three corpora are used, and they differ deliberately.

The \emph{mutation corpus} (Sections~\ref{sec:rq1}--\ref{sec:rq3}) is generated per rule from that rule's own regular expression. It uses $n=120$ samples for each of the 24 analysed rules in each of the 12 embeddings. After inapplicable combinations are excluded, 251 rule--context cells remain populated.

The \emph{comparative corpora} (Section~\ref{sec:rq4}) hold the negatives fixed at 700: 500 lines of real source drawn from the subject, and 200 structurally credential-shaped non-secrets. Only the construction of the 400 positives varies. Those are 40 samples for each of the 10 credential types shared by all three scanners.

The \emph{real-code corpus} (Section~\ref{sec:rq5}) is 292{,}527 lines of unmodified source from four open-source projects. Each is pinned by commit in the replication package, and scanned line by line as added diff lines.

\subsection{Determinism and seeding}

The mutation and comparative harnesses are seeded (\texttt{20260823}) and reproduce identically; we verified this directly by re-running each and diffing every recorded field except wall-clock time, and \texttt{verify\_paper.py} in the replication package re-derives every numeric claim in this manuscript from the raw output on every reproduction run. The cross-tool harness is \emph{not} seeded: it draws credential material from \texttt{secrets.SystemRandom}. We report this rather than retrofit a seed. The property it establishes is that a rule fails on \emph{every} hyphen-terminated credential, not on particular ones. That is a claim about the whole class, and fixing one draw would not strengthen it. Its conclusions were stable across runs, and the per-type outcomes it produces are deterministic in the sense established in Section~\ref{sec:paired}: every credential type yields a rate of exactly $0.000$ or exactly $1.000$.

\subsection{Statistical treatment}

Proportions are reported with Wilson 95\% confidence intervals. These behave correctly at the boundary values of 0 and 1, which occur throughout this data. Between-tool comparisons are \emph{not} tested at the sample level. Per-type detection is deterministic, so samples within a credential type are repetitions rather than independent observations. The effective unit of analysis is therefore the credential type. Section~\ref{sec:paired} sets out the consequences and reports an exact paired sign test over the ten types. We report effect sizes and intervals in preference to $p$-values throughout, and where we do test, we report the discordant cases themselves rather than a statistic alone.

\section{RQ1: Context Sensitivity}
\label{sec:rq1}

\begin{figure*}[t]
\centering
\includegraphics[width=0.92\textwidth]{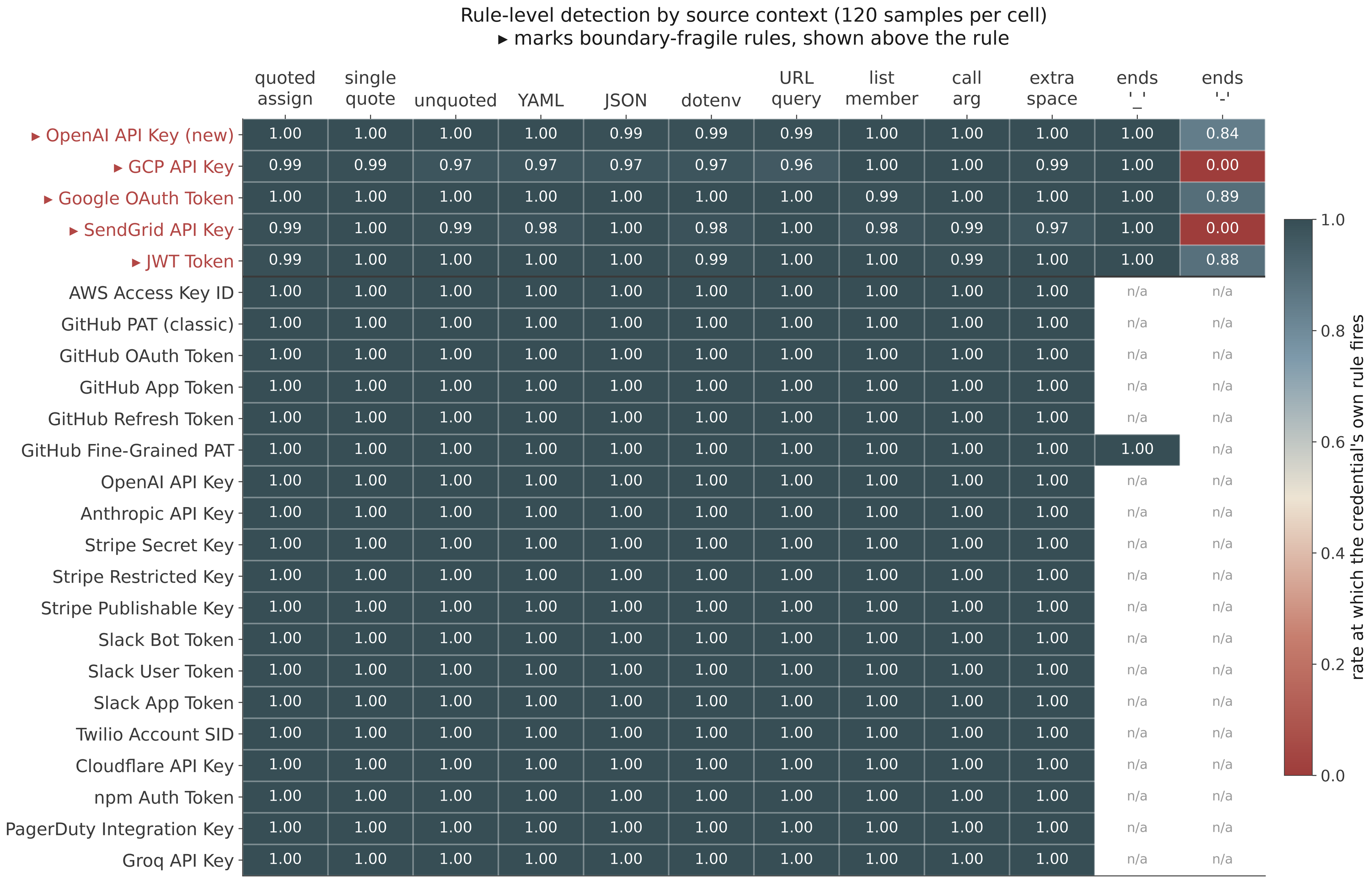}
\caption{Rule-level detection across 24 rules and 12 contexts, 120 samples per cell. Blue indicates the credential's own rule fired; red indicates it did not. ``n/a'' marks embeddings not applicable to that rule. A leading $\blacktriangleright$ marks a boundary-fragile rule, listed above the robust rules and coloured red.}
\label{fig:heatmap}
\end{figure*}

Fig.~\ref{fig:heatmap} gives the full rule $\times$ context matrix and Table~\ref{tab:contexts} the aggregate, both from a seeded run (\texttt{20260823}) that reproduces exactly; we confirmed this by re-running the harness and diffing every field but wall-clock time. Across ten of twelve contexts the rule-level detection rate never falls below $0.9976$. Forcing the final character to an underscore is likewise harmless ($1.0000$). One context breaks it: when the final character is a hyphen --- which the affected rules' own character classes permit --- the aggregate falls to $\mathbf{0.5233}$, with $23.2\%$ of samples rescued only by the entropy fallback and $\mathbf{24.5\%}$ reported by nothing at all.

\begin{table}[t]
\centering
\caption{Aggregate outcome by context (24 rules, $n{=}120$ per cell)}
\label{tab:contexts}
\small
\begin{tabular}{@{}lrrrr@{}}
\toprule
Context & $n$ & Rule fires & Fallback & Missed \\
\midrule
quoted assignment & 2880 & 0.9990 & 0.0003 & 0.0007 \\
single-quoted & 2880 & 0.9997 & 0.0003 & 0.0000 \\
unquoted & 2880 & 0.9986 & 0.0000 & 0.0014 \\
YAML value & 2880 & 0.9983 & 0.0000 & 0.0017 \\
JSON value & 2880 & 0.9986 & 0.0014 & 0.0000 \\
dotenv & 2880 & 0.9976 & 0.0000 & 0.0024 \\
URL query & 2880 & 0.9979 & 0.0000 & 0.0021 \\
list member & 2880 & 0.9990 & 0.0000 & 0.0010 \\
call argument & 2880 & 0.9993 & 0.0000 & 0.0007 \\
extra whitespace & 2880 & 0.9986 & 0.0003 & 0.0010 \\
ends in \texttt{\_} & 720 & 1.0000 & 0.0000 & 0.0000 \\
\textbf{ends in \texttt{-}} & \textbf{600} & \textbf{0.5233} & \textbf{0.2317} & \textbf{0.2450} \\
\bottomrule
\end{tabular}
\end{table}

\subsection{Reproducibility and repeated-run robustness}
\label{sec:reprod-rq1}

Two of the five fragile rules are deterministic in the strongest sense available to this method: \emph{GCP API Key} and \emph{SendGrid API Key} score $BR=0.0000$ in \emph{every one} of ten independently seeded repetitions ($n=120$ each), with zero variance. Both are governed by fixed-count quantifiers (\texttt{\{35\}}, \texttt{\{43\}}), and their failure depends only on whether the credential ends in a hyphen, not on any other property of the generated string.

The remaining three fragile rules --- governed by variable-count quantifiers (\texttt{\{50,\}}, \texttt{\{68,\}}, \texttt{\{10,\}}) --- do not share this property. Across the same ten repetitions their truncated-match rate ranges from $92.5\%$ to $98.3\%$ (pooled mean $95.3\%$), compared with $84$--$89\%$ in the single reproducible run reported in Table~\ref{tab:contexts} and Table~\ref{tab:metrics}. This is not measurement noise in the ordinary sense: \texttt{exrex} generates a variable-count quantifier at a random length above its stated minimum, so whether the regex engine's backtracking finds a valid truncated match depends on how much slack that length gives it and on the character immediately before the truncation point. A fixed-count rule offers no such slack, which is precisely why those two rules are exactly deterministic and these three are not. We report the single seeded run as the paper's reproducible reference value throughout, and this range as evidence that the qualitative finding --- severe, mechanism-specific degradation, an order of magnitude below the $\geq 0.9976$ baseline --- does not depend on which draw was reported.

\subsection{The three metrics, per rule}

Table~\ref{tab:metrics} reports $BR$, $RLP$ and $SP$ for the five rules on which they are not all $1.0000$; the remaining nineteen rules score $1.0000$ on all three. The table makes two points that the aggregate in Table~\ref{tab:contexts} cannot.

First, $BR$ and the preservation metrics are orthogonal, and the two catastrophic rules fail in opposite ways. SendGrid and GCP both have $BR = 0.0000$: neither rule ever fires on a hyphen-terminated credential, in any of the ten independently seeded repetitions of Section~\ref{sec:reprod-rq1}. But SendGrid has $RLP = 1.0000$ and GCP has $RLP = 0.9116$. SendGrid's failures are \emph{total misses}, so they leave the preservation denominator entirely --- of the credentials it did detect, every one was labelled correctly. GCP's failures are \emph{rescued and downgraded}, so they stay in the denominator and depress the ratio. A single scalar cannot express both; two are needed, and which one moves tells the maintainer whether the fix is a detection problem or a labelling problem.

Second, $SP$ and $RLP$ are numerically identical here, for a reason worth stating explicitly rather than leaving as a coincidence. The \textsc{other} outcome --- a different named rule firing --- has rate exactly $0.0000$ in all 251 recorded cells, so the only substitute label available is the entropy fallback, whose severity is fixed at \textsc{medium}. A fallback rescue therefore preserves severity only for a rule that is itself \textsc{medium}, and no \textsc{medium} rule in this table is ever rescued. The two metrics would separate on a scanner with overlapping rules of equal severity; on this one they cannot. We report both because the distinction is a property of the metric definitions, not of this dataset, and a replication on another tool should not have to rediscover it.

\begin{table}[t]
\centering
\caption{Boundary robustness, rule-label preservation and severity preservation. Only the five rules scoring below $1.0000$ are shown; the other nineteen score $1.0000$ throughout.}
\label{tab:metrics}
\small
\setlength{\tabcolsep}{4pt}
\begin{tabular}{@{}llrrr@{}}
\toprule
Rule & Severity & $BR$ & $RLP$ & $SP$ \\
\midrule
SendGrid API Key & \textsc{critical} & 0.0000 & 1.0000 & 1.0000 \\
GCP API Key & \textsc{high} & 0.0000 & 0.9116 & 0.9116 \\
OpenAI API Key (new) & \textsc{critical} & 0.8417 & 0.9861 & 0.9861 \\
JWT Token & \textsc{high} & 0.8833 & 1.0000 & 1.0000 \\
Google OAuth Token & \textsc{high} & 0.8917 & 1.0000 & 1.0000 \\
\midrule
\multicolumn{2}{@{}l}{macro mean, all 24 rules} & 0.9007 & 0.9957 & 0.9957 \\
\bottomrule
\end{tabular}
\end{table}

\section{RQ2: Mechanism and Marginal Probability}
\label{sec:rq2}

Five rules are boundary-fragile: \emph{OpenAI API Key (new)}, \emph{GCP API Key}, \emph{Google OAuth Token}, \emph{SendGrid API Key}, and \emph{JWT Token}.

\textbf{Mechanism 1: the anchor.} Each ends with \verb|\b| immediately after a character class containing a hyphen, e.g.
\begin{center}
\lstinline!r"\bSG\.[a-zA-Z0-9_-]{22}\.[a-zA-Z0-9_-]{43}\b"!
\end{center}
A hyphen is not a word character. When the final generated character is a hyphen and the following text character is also non-word --- a closing quote, in the canonical case --- no word boundary exists there and the match fails before any suppression logic runs. We confirmed that in failing samples every suppression predicate returns \texttt{False}: the scanner's own heuristics judge the value to be a credential, yet nothing is reported.

\begin{figure*}[t]
\centering
\includegraphics[width=0.92\textwidth]{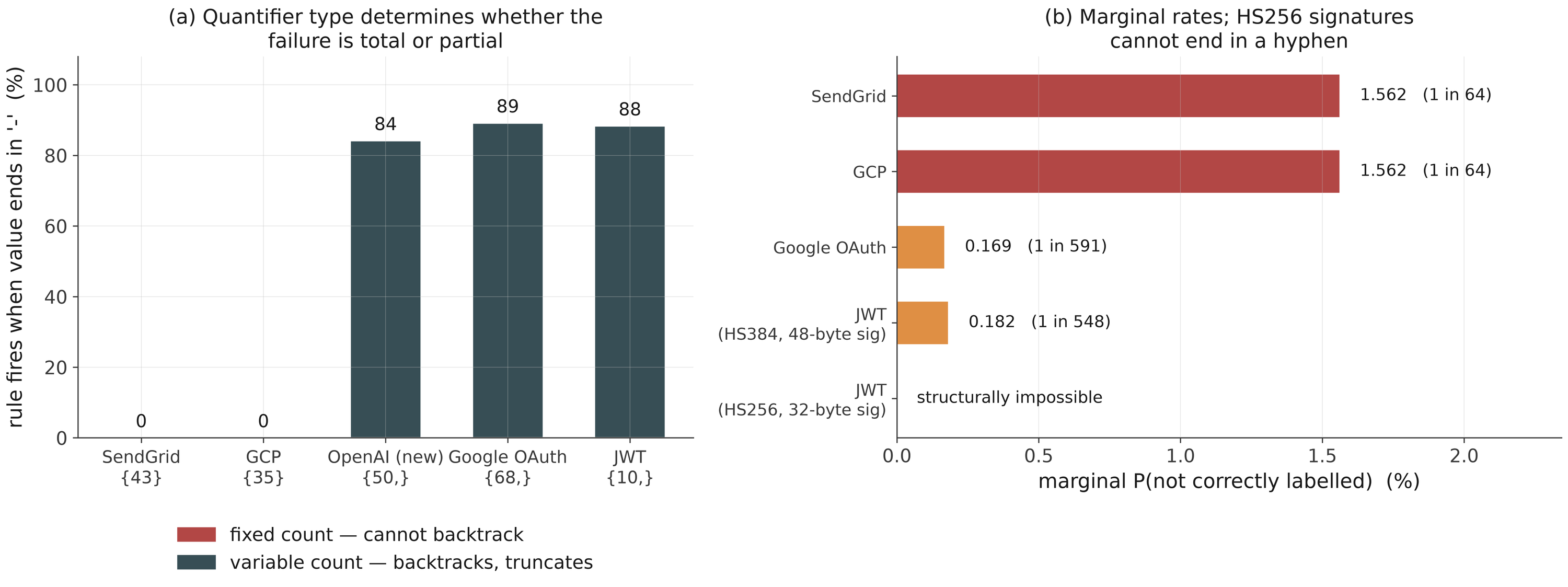}
\caption{(a) Quantifier type determines whether the failure is total or partial. (b) Corrected marginal rates; the encoding structure of HS256 signatures makes the defect unreachable for them.}
\label{fig:mechanism}
\end{figure*}

\textbf{Mechanism 2: quantifier type.} The failure is not uniform, and the split is exact (Fig.~\ref{fig:mechanism}a): fixed-count quantifiers admit no backtracking and fail totally, while variable-count quantifiers backtrack and match one character less.

\begin{center}\small
\begin{tabular}{@{}llr@{}}
\toprule
Rule & Quantifier & Rule fires \\
\midrule
SendGrid API Key & \texttt{\{43\}} fixed & 0.0\% \\
GCP API Key & \texttt{\{35\}} fixed & 0.0\% \\
OpenAI API Key (new) & \texttt{\{50,\}} variable & 84.2\% \\
JWT Token & \texttt{\{10,\}} variable & 88.3\% \\
Google OAuth Token & \texttt{\{68,\}} variable & 89.2\% \\
\bottomrule
\end{tabular}
\end{center}

The variable case is not benign: the rule reports a \emph{truncated} credential. We isolated this on a minimal pair --- \verb|\bX[a-zA-Z0-9_-]{10}\b| fails on a hyphen-terminated value while \verb|\bX[a-zA-Z0-9_-]{5,}\b| succeeds by matching one character fewer.

\begin{figure*}[t]
\centering
\includegraphics[width=0.92\textwidth]{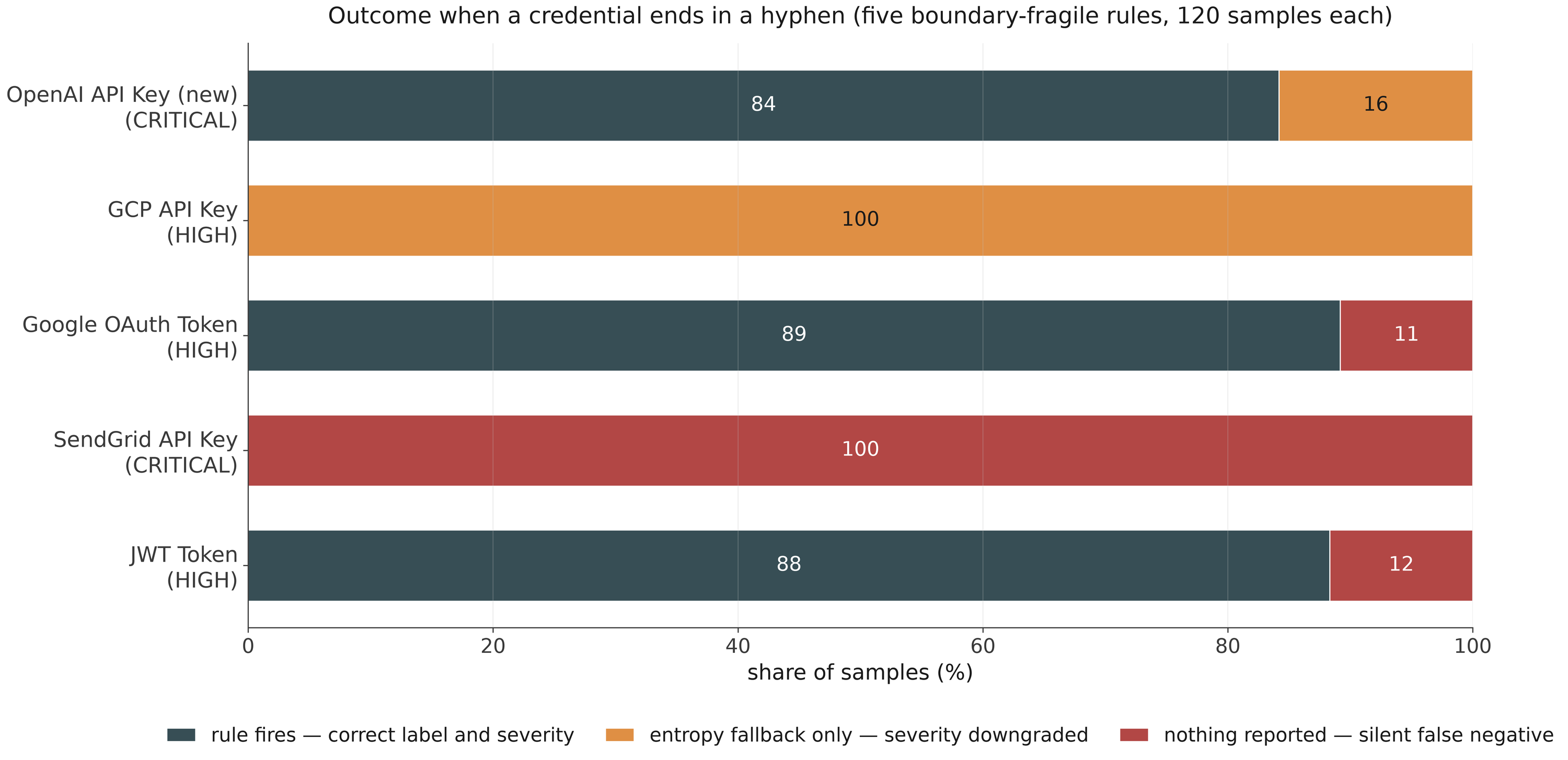}
\caption{Outcome decomposition for the five boundary-fragile rules when the credential ends in a hyphen. The two rules with $BR = 0$ fail in opposite ways: SendGrid is never reported at all, while GCP is always reported but always downgraded. A tool-level metric would score GCP and OpenAI as fully detected.}
\label{fig:fragile}
\end{figure*}

\textbf{Mechanism 3: why some failures are silent.} The entropy fallback captures candidates with \lstinline!['\"]([a-zA-Z0-9+/=_\-]{20,})['\"]!, a class excluding the dot. Credentials containing a dot (SendGrid, Google OAuth, JWT) cannot be rescued and become silent misses; dotless ones (GCP, OpenAI) are rescued but relabelled from \textsc{critical}/\textsc{high} to \textsc{medium} (Fig.~\ref{fig:fragile}). This is the distinction $RLP(\rho)$ and $SP(\rho)$ are defined to capture, and it is why GCP's preservation score is the lowest in Table~\ref{tab:metrics} despite its detection failures being the \emph{least} silent of the five.

\subsection{Marginal, not conditional, probability}

The rates above are \emph{conditional} on the credential ending in a hyphen. The marginal rate is
\[
P(\text{fail}) = P(\text{ends in ``-''}) \times P(\text{fail} \mid \text{ends in ``-''}),
\]
and $P(\text{ends in ``-''})$ is \emph{not} $1/64$ for every type. Two cases must be separated.

For types whose tail is a uniform draw over the 64-symbol base64url-style alphabet (SendGrid, GCP, OpenAI, Google OAuth), $P(\text{ends in ``-''}) = 1/64$ exactly, since the alphabet contains exactly one hyphen symbol and every symbol is equally likely; we report this exact value rather than a simulated one, and confirm it is not being misapplied with five independent $2\times10^5$-draw Monte Carlo checks from a cryptographic generator ($0.015555$--$0.015825$ across the five checks, consistent with $1/64 = 0.015625$ to within sampling error).

For JSON Web Tokens the tail is a base64url \emph{encoding} of a signature of $L$ bytes, and the final character carries only the residual bits: 6 bits when $L \equiv 0 \pmod 3$, 4 bits when $L \equiv 2$, and 2 bits when $L \equiv 1$. The character ``-'' has base64url index 62, which is not a multiple of 4 or 16, so it is reachable \emph{only} when $L \equiv 0 \pmod 3$. HS256 uses a 32-byte signature ($32 \bmod 3 = 2$), and we confirm empirically that only 16 distinct final characters occur and ``-'' never does. \textbf{An HS256 JWT is therefore structurally immune to this defect.} HS384 (48 bytes, $\equiv 0$) is not.

Table~\ref{tab:marginal} gives the corrected marginal rates. The practically consequential case is SendGrid, at exactly one credential in 64; the JWT case, which an earlier draft of this work overstated, is either impossible or roughly one in 548 depending on the signature algorithm.

\begin{table}[t]
\centering
\caption{Marginal probability that a credential is not correctly labelled}
\label{tab:marginal}
\small
\setlength{\tabcolsep}{3.5pt}
\begin{tabular}{@{}lrrl@{}}
\toprule
Credential type & $P(\text{``-''})$ & Marginal & Frequency \\
\midrule
SendGrid API Key & 1/64 & 0.015625 & 1 in 64 (silent miss) \\
GCP API Key & 1/64 & 0.015625 & 1 in 64 (downgraded) \\
Google OAuth Token & 1/64 & 0.001692 & 1 in 591 \\
JWT, HS384 (48 B) & 1/64 & 0.001823 & 1 in 548 \\
JWT, HS256 (32 B) & 0 & 0.000000 & impossible \\
\bottomrule
\end{tabular}
\end{table}

The remedy is small and local: replace the trailing \verb|\b| with a negative lookahead, \verb+(?![A-Za-z0-9_])+, in the five affected rules. We do not merely propose this --- Section~\ref{sec:repair} applies it, re-runs the full battery, and reports a regression we found in a plausible first attempt at the same fix.

\section{Automated Repair and Fix Validation}
\label{sec:repair}

Diagnosing a defect and proposing a remedy is not evidence that the remedy works. This section applies two candidate repairs to the five affected rules, in memory, without modifying the subject repository, and re-runs the full boundary-mutation battery against each.

\subsection{Two candidates}

Both replace the trailing \verb|\b| that Section~\ref{sec:rq2} identifies as the fault:
\begin{itemize}
\item \textbf{Candidate A}, \verb+(?=[^\w-]|$)+, our first attempt: require the following character to be neither a word character nor a hyphen.
\item \textbf{Candidate B}, \verb+(?![A-Za-z0-9_])+, the recommended fix: require only that the following character is not a word character, which is what \verb|\b| itself requires and is the minimal change that removes the hyphen special case.
\end{itemize}
Both are applied to exactly the five rules of Table~\ref{tab:metrics}, holding every other rule and the entropy fallback unchanged.

\subsection{Both candidates repair the defect}

Fig.~\ref{fig:repair}a re-runs the full boundary-mutation battery under each candidate. Both restore $BR = 1.0$ on all five rules: the hyphen-terminated case that previously failed at $0$--$89\%$ now fires at $100\%$ under either fix. Neither candidate introduces a single new finding on the 292{,}527-line real-code corpus of Section~\ref{sec:rq5}: 67 findings under the baseline rule table, and 67 under each candidate, file for file.

\subsection{Candidate A regresses; Candidate B does not}
\label{sec:regression}

A repair to an anchoring assertion can fail in a direction the boundary battery does not by itself probe: it can become \emph{stricter} than the assertion it replaces, silently rejecting inputs the original correctly accepted. We therefore added one further probe, run under each candidate: a credential whose \emph{own} final character is an ordinary word character, immediately followed by a hyphen in the surrounding text (for example, a credential embedded as \verb|cfg_value = "...X-suffix"|). The original \verb|\b| accepts this case --- the boundary between a word character and a hyphen is a valid word boundary --- so a correct repair must continue to accept it.

\begin{figure*}[t]
\centering
\includegraphics[width=0.92\textwidth]{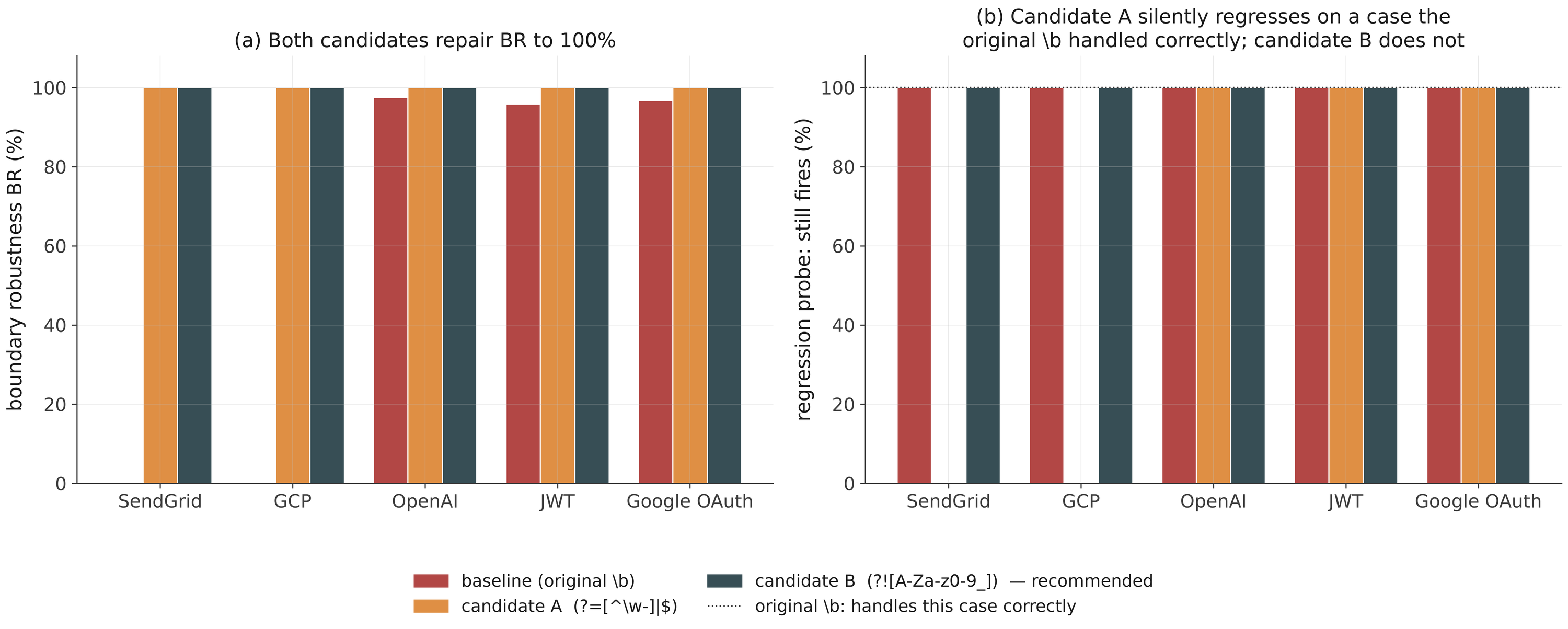}
\caption{Fix validation. (a) Both candidates repair boundary robustness to $1.0$ on all five affected rules. (b) A regression probe --- a credential ending in a word character, followed by a hyphen --- that the original rule handled correctly. Candidate A silently fails it on the two fixed-count-quantifier rules; Candidate B does not.}
\label{fig:repair}
\end{figure*}

Candidate A does not. Fig.~\ref{fig:repair}b shows it regressing to total failure ($0\%$) on exactly the two rules governed by fixed-count quantifiers, SendGrid and GCP, on precisely this case --- the same failure mode Section~\ref{sec:rq2} diagnosed, reintroduced by the fix. The cause is the same construct that caused the original defect: \verb+(?=[^\w-]|$)+ forbids a hyphen in the following text unconditionally, including here, where a hyphen legitimately follows a word-character-terminated credential. Candidate B imposes no such restriction and fires at $100\%$ on all five rules under this probe, matching the original rule's behaviour on the case it already handled correctly.

This is the paper's second methodological point, not only its first result: a plausible one-line regex repair, chosen by inspection and applied to exactly the rules it was meant to fix, introduced a new defect of the same kind it removed. We found this only because we ran an oracle-based probe against the fix rather than reading it. We recommend Candidate B, \verb+(?![A-Za-z0-9_])+, and report both results because a method for finding hidden assumptions in a detection rule should be applied to its own proposed repair, not only to the original.

\section{RQ3: Does the Method Transfer?}
\label{sec:rq3}

We applied the identical battery to Gitleaks and TruffleHog, over 4{,}320 files. For external tools the expected rule per credential type is determined \emph{empirically} from each tool's own behaviour in the canonical context --- whichever \texttt{RuleID} or \texttt{DetectorName} it fires there --- so no tool is judged against another project's naming. Types a tool never fires on are excluded from its aggregate and reported separately, so coverage gaps are not scored as fragility.

\begin{figure*}[t]
\centering
\includegraphics[width=0.92\textwidth]{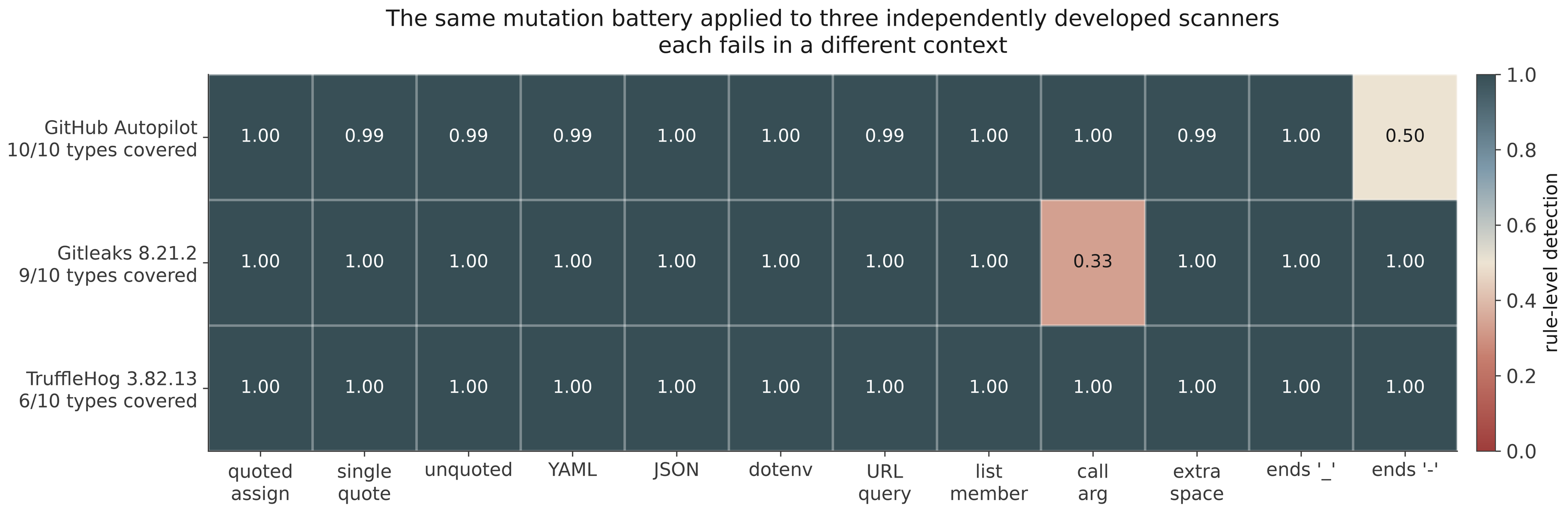}
\caption{The same mutation battery applied to three independently developed scanners. The battery reveals scanner-specific degradation patterns; no tested context degraded all three scanners.}
\label{fig:crosstool}
\end{figure*}

Fig.~\ref{fig:crosstool} shows the result: \textbf{the method transfers, and exposes a distinct behavioural profile in each tool}. Two of the three degrade, and they degrade in different contexts; the third does not degrade at all, which is itself a finding the same battery establishes.

\textbf{A second, unrelated defect in Gitleaks.} Gitleaks is unaffected by the trailing hyphen, but its rule-level detection falls to $0.333$ when a credential appears as a bare call argument. Per-type inspection shows this is not uniform: \texttt{github-pat}, \texttt{github-oauth} and \texttt{slack-bot-token} are unaffected, while \texttt{aws-access-token}, \texttt{gcp-api-key}, \texttt{npm-access-token}, \texttt{stripe-access-token}, \texttt{sendgrid-api-token} and \texttt{jwt} all drop to $0.000$ --- total misses.

\begin{figure*}[t]
\centering
\includegraphics[width=0.92\textwidth]{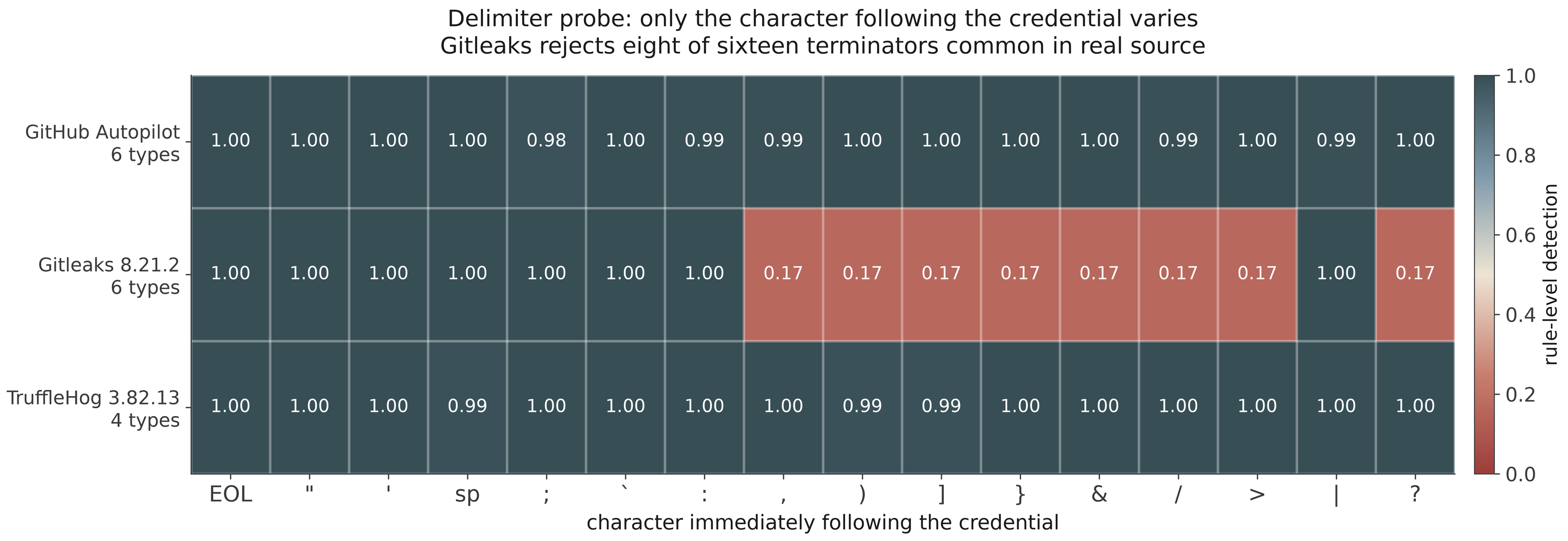}
\caption{Delimiter probe: only the character immediately following the credential varies. Gitleaks rejects eight of sixteen terminators that occur routinely in real source.}
\label{fig:delim}
\end{figure*}

To isolate the mechanism we ran a delimiter probe varying \emph{only} the single character following the credential (Fig.~\ref{fig:delim}). Gitleaks accepts end-of-line, both quote characters, whitespace, semicolon, backtick and colon, and rejects comma, closing parenthesis, closing bracket, closing brace, ampersand, forward slash, angle bracket and question mark. We confirmed the mechanism directly from source: Gitleaks' public configuration, \texttt{config/gitleaks.toml} at tag \texttt{v8.21.2}, defines each affected rule (\texttt{aws-access-token}, \texttt{gcp-api-key}, \texttt{npm-access-token}, \texttt{sendgrid-api-token}, \texttt{stripe-access-token}, \texttt{jwt}) with an identical trailing non-capturing group appended to the credential-capturing group, \lstinline!(?:['|\"|\n|\r|\s|\x60|;]|$)!, a Go regular-expression character class containing only the literal characters \texttt{'}, \texttt{|}, \texttt{"}, newline, carriage return, whitespace, backtick, and semicolon, or end-of-string. Every terminator we found rejected is provably absent from this class; every terminator we found accepted is provably present in it. The three unaffected types --- \texttt{github-pat}, \texttt{github-oauth}, \texttt{slack-bot-token} --- carry no such trailing group in their source regex at all, which is the source-level reason our data found them robust. We additionally verified the effect on minimal hand-written cases outside the harness: \texttt{cfg = "AKIA\ldots"} is detected, \texttt{f(AKIA\ldots)} is not.

The security relevance is direct. Every rejected character occurs routinely where credentials appear: \texttt{f(token)} in code, \texttt{[tok, x]} in a list, and --- most consequentially --- \texttt{?key=SECRET\&next=1} in a URL, which is among the more common shapes in which credentials actually leak.

\textbf{TruffleHog shows no boundary fragility.} It is robust across every context and terminator, on the types it covers --- but it covers the fewest (6 of 10 canonically). Its detectors are co-occurrence- and verification-oriented rather than shape-oriented, which is exactly the trade-off Section~\ref{sec:rq4} quantifies.

\textbf{Taxonomy.} Three boundary-handling strategies, three failure profiles: an \emph{implicit} anchor (\verb|\b|) yields probabilistic failure depending on the credential's own final character; an \emph{explicit} terminator allowlist yields deterministic failure depending on the following character; and \emph{no} boundary dependence yields robustness at the cost of coverage. The method exposes all three; the defect class is tool-specific, the vulnerability to boundary assumptions is not.

\section{RQ4: Comparison, Corpus, and Deployment Condition}
\label{sec:rq4}

\begin{table*}[t]
\centering
\caption{Three scanners on identical corpora (400 positives, 700 negatives), with Wilson 95\% intervals. No value is emphasised: Section~\ref{sec:paired} shows that none of the between-tool differences is statistically significant at the correct unit of analysis.}
\label{tab:comparative}
\footnotesize
\setlength{\tabcolsep}{3pt}
\begin{tabular}{@{}llllll@{}}
\toprule
Corpus & Tool & Precision [95\% CI] & Recall [95\% CI] & F1 & TP / FP / FN \\
\midrule
\multirow{3}{*}{A: format-spec}
 & GitHub Autopilot & 0.959 [0.936, 0.974] & 1.000 [0.991, 1.000] & 0.979 & 400 / 17 / 0 \\
 & Gitleaks 8.21.2 & 1.000 [0.988, 1.000] & 0.800 [0.758, 0.836] & 0.889 & 320 / 0 / 80 \\
 & TruffleHog 3.82.13 & 1.000 [0.984, 1.000] & 0.600 [0.551, 0.647] & 0.750 & 240 / 0 / 160 \\
\midrule
\multirow{3}{*}{B: structural}
 & GitHub Autopilot & 0.959 [0.936, 0.974] & 1.000 [0.991, 1.000] & 0.979 & 400 / 17 / 0 \\
 & Gitleaks 8.21.2 & 1.000 [0.989, 1.000] & 0.900 [0.867, 0.926] & 0.947 & 360 / 0 / 40 \\
 & TruffleHog 3.82.13 & 1.000 [0.984, 1.000] & 0.600 [0.551, 0.647] & 0.750 & 240 / 0 / 160 \\
\bottomrule
\end{tabular}
\end{table*}

Corpus construction alone moves Gitleaks' recall from $0.800$ to $0.900$: its JWT rule validates internal token structure and correctly refuses a ``JWT'' whose segments are random characters. The subject's scanner is unmoved at $1.000$ --- not because it is more capable, but because its rules are shallower (prefix, class, length). That shallowness also produces its 17 false positives where the others have none.

\subsection{The comparison supports no ranking}
\label{sec:paired}

The intervals in Table~\ref{tab:comparative} are computed over samples, and taken at face value they do not overlap. That would be the wrong conclusion to draw, for a reason visible in the raw data: \emph{per-type detection is deterministic}. Across both corpora and all three tools, every one of the sixty per-type recall values is exactly $0.000$ or exactly $1.000$. A tool either detects a credential type or it does not; the 40 samples within a type are 40 repetitions of one outcome, not 40 independent observations.

The effective unit of analysis is therefore the credential \emph{type}, of which there are ten, and a sample-level test would overstate the evidence by a factor of 40. We accordingly compare tools with an exact paired sign test over the ten types, reporting the discordant types themselves rather than a test statistic alone (Table~\ref{tab:paired}).

\begin{table}[t]
\centering
\caption{Exact paired sign test over $n{=}10$ credential types. ``Discordant'' counts types detected by the first tool and not the second, and conversely.}
\label{tab:paired}
\small
\setlength{\tabcolsep}{4pt}
\begin{tabular}{@{}llrr@{}}
\toprule
Corpus & Comparison & Discordant & $p$ \\
\midrule
\multirow{3}{*}{A}
 & Autopilot vs.\ Gitleaks & 2 / 0 & 0.500 \\
 & Autopilot vs.\ TruffleHog & 4 / 0 & 0.125 \\
 & Gitleaks vs.\ TruffleHog & 2 / 0 & 0.500 \\
\midrule
\multirow{3}{*}{B}
 & Autopilot vs.\ Gitleaks & 1 / 0 & 1.000 \\
 & Autopilot vs.\ TruffleHog & 4 / 0 & 0.125 \\
 & Gitleaks vs.\ TruffleHog & 3 / 0 & 0.250 \\
\bottomrule
\end{tabular}
\end{table}

\textbf{No comparison reaches significance at $\alpha=0.05$}, and none comes close. The differences are entirely one-directional --- every discordant type favours the same tool, never the reverse --- which is consistent with a real ordering, but with ten types the sign test cannot resolve one: even a perfect $4/0$ split has $p = 0.125$, because $2^{-4}$ doubled is the smallest two-sided $p$-value ten paired types can produce at that margin. The design is underpowered for ranking by construction, and no larger sample of \emph{credentials} would fix it; only more credential \emph{types} would.

We therefore report Table~\ref{tab:comparative} as a characterisation of where the tools differ and why, and we draw no ranking from it. This bears directly on the competing interest declared in Section~\ref{sec:bias}: the raw numbers favour the author's own tool, and the correct statistical treatment does not support that advantage. Readers should treat the coverage and trade-off structure below, not the recall ordering, as this section's result.

\begin{figure*}[t]
\centering
\includegraphics[width=0.92\textwidth]{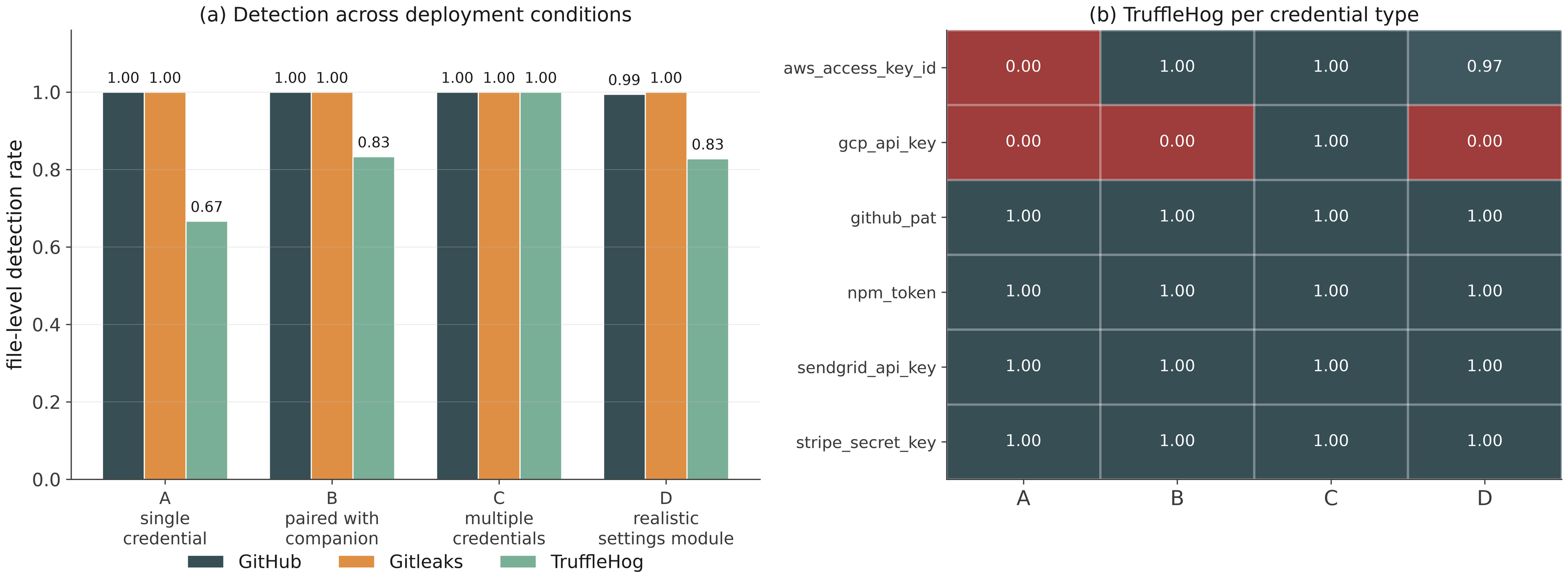}
\caption{Detection across four deployment conditions. TruffleHog's recall depends strongly on corroborating context; the other two tools do not.}
\label{fig:cooc}
\end{figure*}

\textbf{Deployment condition matters more.} Our single-credential-per-file design is adversarial to co-occurrence-based detection, so we evaluated four conditions (Fig.~\ref{fig:cooc}): (A) a credential alone; (B) paired with its companion secret; (C) several credentials in one file; (D) a realistic settings module with imports, comments and unrelated configuration.

TruffleHog's detection rises from $0.667$ (A) to $0.833$ (B) to $1.000$ (C), settling at $0.828$ in the realistic condition (D). Per type, AWS goes from $0.000$ in isolation to $1.000$ once a secret access key accompanies it. Its apparent recall deficit in Table~\ref{tab:comparative} is therefore \textbf{substantially an artefact of the corpus}, not a capability limit, and the single-credential numbers should be read as a lower bound under conditions adversarial to that architecture. The other two tools are near-invariant across conditions.

\section{RQ5: False Positives on the Evaluated Real-Code Corpus}
\label{sec:rq5}

We scanned 292{,}527 lines of unmodified source from four open-source projects --- \texttt{requests}, \texttt{flask}, \texttt{redis-py}, and the subject --- line by line as added diff lines, mirroring the subject's push handler. All four are pinned by commit in the replication package.

This corpus is small and Python-centric, and it is not a sample of anything. We therefore make no claim about real-world false-positive rates in general; the claim is about \emph{these four projects}, and its purpose is to test whether a conclusion drawn from the synthetic corpus survives contact with unmodified code. It does not.

\begin{figure*}[t]
\centering
\includegraphics[width=0.92\textwidth]{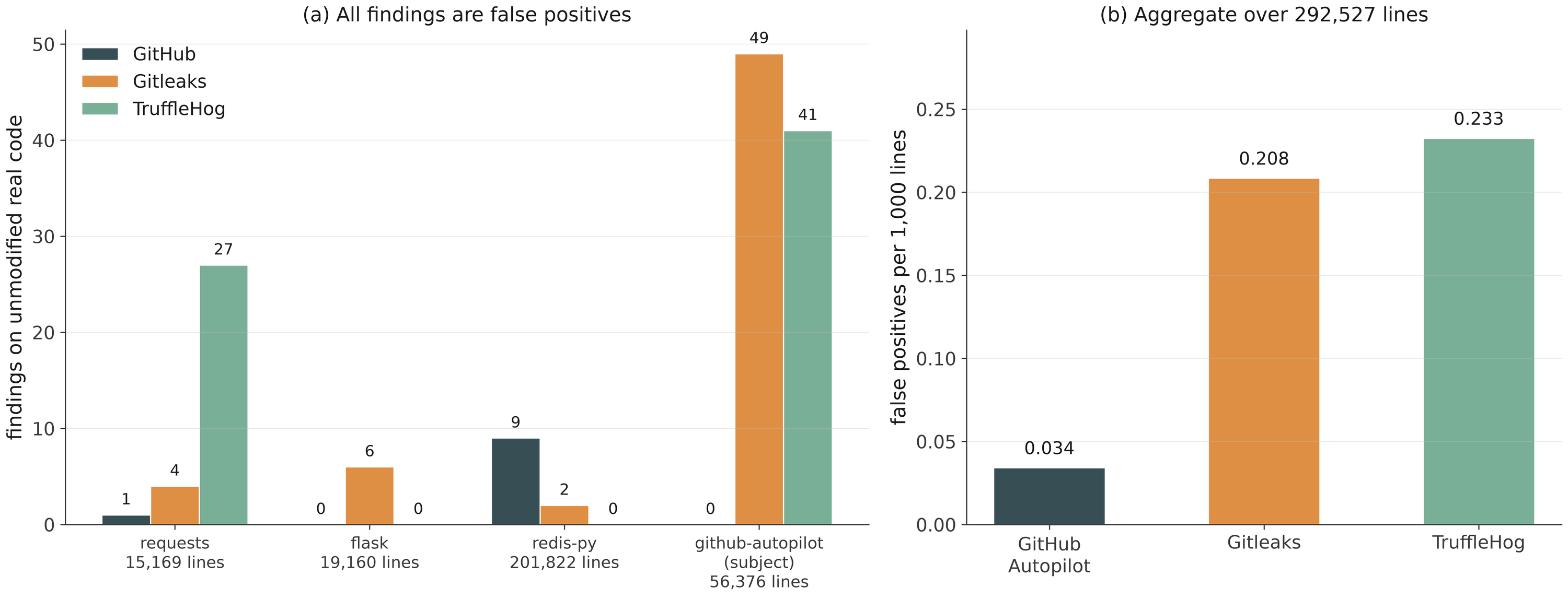}
\caption{False positives on the four unmodified open-source projects evaluated. The ordering inverts relative to the synthetic corpus.}
\label{fig:rw}
\end{figure*}

The ordering \emph{inverts} relative to the synthetic corpus (Fig.~\ref{fig:rw}). On constructed negatives the subject's scanner had the worst precision; on this corpus it has the lowest rate: $0.034$ findings per kLOC against Gitleaks' $0.208$ and TruffleHog's $0.233$. The synthetic corpus over-represented bare high-entropy base64 assignments, to which its unanchored fallback is uniquely vulnerable, and under-represented the documentation and configuration idioms that dominate real repositories.

We inspected all ten of its findings; every one is a false positive, in three groups: six occurrences of a documentation URL \emph{template} (\texttt{redis://[[username]:\allowbreak[password]]\allowbreak @localhost:6379/0}) matched at \textsc{critical} severity; three doctest passwords; and one RFC 3986 unreserved-character constant matched by the entropy fallback.

For fairness: 49 of Gitleaks' and 41 of TruffleHog's findings occur in the subject's own repository, which contains credential-shaped strings in its rule table and fixtures and which its scanner suppresses via path rules tuned for that layout. Restricted to the three third-party projects the totals are 10, 12, and 27.

\section{Discussion}

\subsection{Example-based validation has a blind spot with a shape}
\label{sec:discussion-cost}

All five defective rules in the primary subject, and all six in Gitleaks, pass their projects' own test suites, because every fixture places the credential in a canonical position. The blind spot is not random: it is exactly the set of behaviours that depend on the credential's boundary, and it is invisible to any suite whose fixtures share one embedding. Neither defect we found is exotic --- one fires on roughly one SendGrid key in 63, the other on any credential followed by a comma. The battery that finds both is small and cheap: twelve embeddings generated from the rules themselves, running in under twenty seconds over 24 rules. The cost of \emph{not} running it is a silent false negative in a security control.

\subsection{Rule-level oracles are not optional}

Measuring only whether something was reported would have scored GCP and OpenAI keys as fully detected while their severity was being silently downgraded from \textsc{critical} or \textsc{high} to \textsc{medium}. It would also have concealed our own harness defect (Section~\ref{sec:control}), which a tool-level oracle could not have distinguished from correct behaviour. $RLP(\rho)$ and $SP(\rho)$ exist to make that distinction quantitative, and Table~\ref{tab:metrics} shows why one number is not enough: two rules with identical, maximally bad $BR$ scores differ completely in whether their failures are silent or merely mislabelled, which is the difference between an undetected leak and a deprioritised alert.

\subsection{Boundary strategy is a design axis with real consequences}

The three scanners embody three distinct strategies, and the method surfaces the cost of each. An implicit anchor fails probabilistically, as a function of the credential's own final character. An explicit terminator allowlist fails deterministically, as a function of the character that follows. Boundary-independent detection avoids both failure modes but covers the fewest credential shapes. No scanner dominates across all evaluated dimensions and conditions --- and, as Section~\ref{sec:paired} establishes, our design is not powered to rank them even where it might appear to. What the method delivers is not a winner but a diagnosis: for a given rule table, which of these three failure modes applies, and to which rules.

\subsection{Benchmark construction, not only tool quality, decides what a comparison reports}

Three separate results here are warnings about our own methodology. Corpus construction moves Gitleaks by $+0.10$ recall. Deployment condition moves TruffleHog by $+0.33$ detection. Synthetic versus real negatives invert the precision ordering entirely. A single-corpus, single-condition evaluation of these three tools would have produced a confident conclusion in any of three mutually contradictory directions, depending only on which condition the experimenter happened to pick. We take this as an argument for reporting the sensitivity of a comparison to its own construction as a first-class result, rather than reporting a single table and a ranking.

\section{Threats to Validity}
\label{sec:threats}

\subsection{Construct validity}

Twelve embeddings do not exhaust the boundary space. One family we know is missing is encoding transformations --- base64-wrapped configuration, escaped strings, template interpolation --- which are untested and plausibly hide a further defect family; Section~\ref{sec:future} treats this as the first extension.

Positive corpora are generated rather than harvested, because real leaked credentials cannot be redistributed. As Section~\ref{sec:validity} sets out, this guarantees syntactic and not semantic validity: a defect found on a string the provider would never issue may have no field consequence. Sections~\ref{sec:rq2}--\ref{sec:rq3} do not depend on the distinction, since generation is from the rules under test and the finding is an internal inconsistency in the rule. Section~\ref{sec:rq4} does depend on it, and quantifies how much.

The marginal probabilities in Table~\ref{tab:marginal} assume a uniform final character within each type's alphabet; $P(\text{``-''}) = 1/64$ then follows exactly, and we report that exact value rather than a simulated one, confirming it independently by Monte Carlo rather than by construction alone. Real provider issuance may deviate from uniformity, in which case the rates shift; the JWT rows depend additionally on signature length, which we enumerate rather than assume.

The three variable-quantifier rules in Table~\ref{tab:metrics} carry a construct-validity subtlety the two fixed-quantifier rules do not: their truncated-match rate is itself a random variable, not a fixed property of the rule. \texttt{exrex} generates a length above each quantifier's stated minimum, and whether backtracking finds a valid truncated match depends on how much slack that length gives it. Section~\ref{sec:reprod-rq1} reports both the single reproducible value we use throughout and the range across ten independent repetitions, and explains why only these three rules show it.

Finally, $SP$ requires a severity oracle, and we have one only for the primary subject. Section~\ref{sec:rq3} therefore reports $BR$ and $RLP$ for the external scanners and no $SP$, rather than constructing a severity correspondence the tools do not share.

\subsection{Internal validity}

Three harness defects were found during this work and are reported rather than silently corrected. First, an identifier that triggered the subject's generic keyword rules and masked the defect entirely (Section~\ref{sec:control}); this is documented in full because it is the failure mode a replication is most likely to reproduce. Second, an unsynchronised thread start in the supplementary systems benchmark (Online Resource~1, Section~1). Third, the original mutation harness recorded a fixed seed value in its output metadata without ever calling the corresponding seeding function, so the figures in an earlier draft of this paper were not actually reproducible from that seed despite claiming to be: re-running the unseeded harness a second time reproduced neither the recorded rates nor the recorded \emph{count} of boundary-fragile rules (three rather than five). We added the missing seeding call, confirmed exact reproducibility by re-running the corrected harness and diffing every field but wall-clock time against a second run, and used that genuinely seeded output throughout this paper. We disclose this because a paper that argues for validating claims empirically should not exempt its own infrastructure from the same standard.

The Gitleaks mechanism is confirmed from its published source configuration (Section~\ref{sec:rq3}), quoting the exact character class. The TruffleHog co-occurrence mechanism is behavioural only, established by direct probing rather than by reading its Go source, and we mark it as such. TruffleHog is run with \texttt{--no-verification}; live verification would change its behaviour and was not available in our environment.

The expected-rule freezing procedure of Section~\ref{sec:freeze} cannot detect an attribution error that is both consistent across canonical samples and semantically plausible. We checked the frozen identifiers by inspection but cannot rule this out mechanically.

\subsection{External validity}

The evaluation covers one hardware configuration and one Python version. The real-code corpus is four Python-centric projects pinned by commit, which is not a sample of GitHub and supports no general claim about false-positive rates in the field (Section~\ref{sec:rq5}). Cross-tool results cover ten shared credential types, not any tool's full rule table; extending to full tables would convert our taxonomy into a census, which we have not done.

\subsection{Conclusion validity}

The most important limitation is the one in Section~\ref{sec:paired}: per-type detection is deterministic, so the effective sample size for between-tool comparison is ten credential types, not 400 credentials. At that size no pairwise difference reaches significance, and we draw no ranking. Readers should not read Table~\ref{tab:comparative}'s intervals as licensing one.

TruffleHog's figures in that table are a lower bound under conditions adversarial to its architecture, established by direct probing and quantified in Fig.~\ref{fig:cooc}. Real-code ground truth assumes released public code contains no live credentials; we inspected every finding individually, but an undetected genuine leak would shift the counts.

\section{Future Work}
\label{sec:future}

Five directions follow, ordered by how much they would strengthen the method rather than by effort. First, extend the battery beyond boundary embeddings to encoding transformations, the family we expect most likely to conceal a second defect class. Second, apply the method to the full rule tables of both external scanners rather than ten shared types, converting the taxonomy of Section~\ref{sec:rq3} into a census and, incidentally, supplying the credential types that Section~\ref{sec:paired} shows are the binding constraint on any statistical comparison. Third, enlarge the real-code corpus with an explicit, stated sampling frame across languages, so that false-positive behaviour can be characterised rather than merely observed on four projects. Fourth, obtain severity oracles for the external scanners, which would allow $SP$ to be reported across all three. Fifth, extend the repair-and-validate loop of Section~\ref{sec:repair} to Gitleaks' terminator-allowlist defect, which we have diagnosed but not patched; we would expect the same lesson to recur, that a plausible one-line fix is a hypothesis until it is run against the battery that found the original defect.

\section{Conclusion}

Pattern-based secret scanners are validated by example. Example-based fixtures hold constant the one variable that decides whether an anchored regular expression matches: the text surrounding the credential. Boundary-mutation testing varies it systematically. Credentials are generated from each rule's own expression and embedded in contexts that real credentials occupy. Outcomes are then judged at the rule level rather than the tool level. This yields three metrics, which separate failing to detect a credential from detecting it and labelling it wrongly.

Applied to three independently developed scanners, the method transferred to all three. It exposed a distinct behavioural profile in each. In the first, a hyphen-terminated credential defeats an implicit word-boundary anchor. In the second, eight routine punctuation characters defeat an explicit terminator allowlist, confirmed from source. The third shows no boundary fragility at all, at the cost of the narrowest coverage and a strong dependence on corroborating context. Each defect is reported with its mechanism rather than as a symptom. Each is also reported with the marginal, not merely conditional, probability that it affects a real credential --- an analysis that shows one widely deployed credential format to be structurally immune and another to fail once in 64.

We do not stop at diagnosis. We patch the primary subject's five defective rules, re-run the full battery, and confirm the repair with no regression and no new false positives on real code. We report, alongside that success, a repair candidate that looked equally plausible and was not: an inspection-only fix silently reintroduced the same failure mode on two of the five rules, caught only because we ran the same oracle-based probe against our own patch that we ran against the original defect.

For practice, the implication is twofold: a rule table can be tested against itself, cheaply, finding defects the project's own suite is structurally incapable of observing; and a proposed fix to such a defect is a hypothesis, not a fact, until it is tested the same way. For evaluation, the implication is more cautionary. In three separate places our own benchmark construction could have reversed the conclusion. At the correct unit of analysis, our comparison cannot rank the three tools at all. We report that negative result alongside the positive ones. A method for exposing hidden assumptions in rule-based detection tests, and for validating proposed repairs to them, is worth little if it is not turned on the evaluation that carries it.

\section*{Statements and Declarations}

\textbf{Funding.} This research received no external funding.

\textbf{Competing interests.} The author is the creator and maintainer of GitHub Autopilot, one of the three systems evaluated in this study. To mitigate the resulting researcher bias, the evaluation used a subject version frozen before any measurement, pinned and checksum-verified external tool binaries, byte-identical corpora and execution conditions across all three tools, and retention of all raw outputs including those unfavourable to the subject. No result was filtered after inspection and no defect in the subject was repaired before measurement. Section~\ref{sec:bias} states these controls in full and identifies the specific findings that are unfavourable to the subject, including the paired analysis in Section~\ref{sec:paired}, which withdraws the recall advantage the raw numbers would otherwise give it. The author has no financial interest in, and no relationship with, the maintainers of Gitleaks or TruffleHog.

\textbf{Data availability.} All raw experimental outputs --- per-rule mutation results, cross-tool and delimiter probes, comparative and co-occurrence benchmarks, real-code findings, and the probability analysis --- are included in the replication package as unmodified JSON, together with the redacted finding logs. The complete replication package is archived on Zenodo at \url{https://doi.org/10.5281/zenodo.22114221} \citep{mishra2026zenodo}.

\textbf{Code availability.} All benchmark harnesses, the figure-generation code, a pinned environment specification, and a one-command \texttt{reproduce.sh} are included in the replication package. The subject system is public at \url{https://github.com/Shweta-Mishra-ai/github-autopilot} and pinned at commit \texttt{38b2013}; the comparison tools are official release binaries at versions 8.21.2 and 3.82.13, verified by SHA-256 at reproduction time.

\textbf{Author contributions.} Shweta Mishra is the sole author and is responsible for the study design, implementation, execution, analysis, and manuscript.

\textbf{Ethical approval.} Not applicable. This study involved no human participants and no personal data. All credentials used are synthetic, generated from published format specifications; no live or leaked credential was collected, tested, or redistributed.

\textbf{Replication Package.} The package contains all harnesses, raw JSON results, redacted finding logs, figure code, a pinned environment, and a one-command \texttt{reproduce.sh}, together with the complete raw data behind the operational characterisation of Online Resource~1. The subject is pinned at commit \texttt{38b2013} of \url{https://github.com/Shweta-Mishra-ai/github-autopilot}, the real-code corpora at their scanned commits, and the comparison tools at Gitleaks 8.21.2 and TruffleHog 3.82.13 with recorded SHA-256 digests. An archived, versioned snapshot of the package is deposited on Zenodo at \url{https://doi.org/10.5281/zenodo.22114221} \citep{mishra2026zenodo}. No number in this paper was estimated or extrapolated; each traces to a recorded script output.



\begin{thebibliography}{26}
\providecommand{\natexlab}[1]{#1}
\providecommand{\url}[1]{#1}
\csname url@samestyle\endcsname
\providecommand{\newblock}{\relax}
\providecommand{\bibinfo}[2]{#2}
\providecommand{\BIBentrySTDinterwordspacing}{\spaceskip=0pt\relax}
\providecommand{\BIBentryALTinterwordstretchfactor}{4}
\providecommand{\BIBentryALTinterwordspacing}{\spaceskip=\fontdimen2\font plus
\BIBentryALTinterwordstretchfactor\fontdimen3\font minus
  \fontdimen4\font\relax}
\providecommand{\BIBforeignlanguage}[2]{{%
\expandafter\ifx\csname l@#1\endcsname\relax
\typeout{** WARNING: IEEEtranN.bst: No hyphenation pattern has been}%
\typeout{** loaded for the language `#1'. Using the pattern for}%
\typeout{** the default language instead.}%
\else
\language=\csname l@#1\endcsname
\fi
#2}}
\providecommand{\BIBdecl}{\relax}
\BIBdecl

\bibitem[Meli et~al.(2019)Meli, McNiece, and Reaves]{meli2019}
M.~Meli, M.~R. McNiece, and B.~Reaves, ``How bad can it git? {C}haracterizing
  secret leakage in public {GitHub} repositories,'' in \emph{Proc. NDSS}, 2019.

\bibitem[{Gitleaks contributors}(2024)]{gitleaks}
{Gitleaks contributors}, ``Gitleaks: a {SAST} tool for detecting secrets,
  v8.21.2,'' 2024, \url{https://github.com/gitleaks/gitleaks}.

\bibitem[{Truffle Security Co.}(2024)]{trufflehog}
{Truffle Security Co.}, ``{TruffleHog}, v3.82.13,'' 2024,
  \url{https://github.com/trufflesecurity/trufflehog}.

\bibitem[Saha et~al.(2020)Saha, Denning, Srikumar, and Kasera]{saha2020}
A.~Saha, T.~Denning, V.~Srikumar, and S.~K. Kasera, ``Secrets in source code:
  Reducing false positives using machine learning,'' in \emph{Proc. COMSNETS},
  2020, pp. 168--175.

\bibitem[Rahman et~al.(2019)Rahman, Parnin, and Williams]{rahman2019}
A.~Rahman, C.~Parnin, and L.~Williams, ``The seven sins: Security smells in
  {I}nfrastructure as {C}ode scripts,'' in \emph{Proc. ICSE}, 2019, pp.
  164--175.

\bibitem[Basak et~al.(2022)Basak, Neil, Reaves, and Williams]{basak2022}
S.~K. Basak, L.~Neil, B.~Reaves, and L.~Williams, ``What are the practices for
  secret management in software artifacts?'' in \emph{Proc. IEEE SecDev}, 2022,
  pp. 69--76.

\bibitem[Jia and Harman(2011)]{jia2011}
Y.~Jia and M.~Harman, ``An analysis and survey of the development of mutation
  testing,'' \emph{IEEE Trans. Softw. Eng.}, vol.~37, no.~5, pp. 649--678,
  2011.

\bibitem[DeMillo et~al.(1978)DeMillo, Lipton, and Sayward]{demillo1978}
R.~A. DeMillo, R.~J. Lipton, and F.~G. Sayward, ``Hints on test data selection:
  Help for the practicing programmer,'' \emph{Computer}, vol.~11, no.~4, pp.
  34--41, 1978.

\bibitem[Chen et~al.(2018)Chen, Kuo, Liu, Poon, Towey, Tse, and
  Zhou]{chen2018metamorphic}
T.~Y. Chen, F.-C. Kuo, H.~Liu, P.-L. Poon, D.~Towey, T.~H. Tse, and Z.~Q. Zhou,
  ``Metamorphic testing: A review of challenges and opportunities,'' \emph{ACM
  Comput. Surv.}, vol.~51, no.~1, pp. 1--27, 2018.

\bibitem[Segura et~al.(2016)Segura, Fraser, Sanchez, and
  Ruiz-Cort{\'e}s]{segura2016}
S.~Segura, G.~Fraser, A.~B. Sanchez, and A.~Ruiz-Cort{\'e}s, ``A survey on
  metamorphic testing,'' \emph{IEEE Trans. Softw. Eng.}, vol.~42, no.~9, pp.
  805--824, 2016.

\bibitem[Godefroid et~al.(2008)Godefroid, Kiezun, and Levin]{godefroid2008}
P.~Godefroid, A.~Kiezun, and M.~Y. Levin, ``Grammar-based whitebox fuzzing,''
  in \emph{Proc. PLDI}, 2008, pp. 206--215.

\bibitem[Claessen and Hughes(2000)]{claessen2000}
K.~Claessen and J.~Hughes, ``{QuickCheck}: A lightweight tool for random
  testing of {H}askell programs,'' in \emph{Proc. ICFP}, 2000, pp. 268--279.

\bibitem[Chapman and Stolee(2016)]{chapman2016}
C.~Chapman and K.~T. Stolee, ``Exploring regular expression usage and context
  in {P}ython,'' in \emph{Proc. ISSTA}, 2016, pp. 282--293.

\bibitem[Michael et~al.(2019)Michael, Donohue, Davis, Lee, and
  Servant]{michael2019}
L.~G. Michael, J.~Donohue, J.~C. Davis, D.~Lee, and F.~Servant, ``Regexes are
  hard: Decision-making, difficulties, and risks in programming regular
  expressions,'' in \emph{Proc. ASE}, 2019, pp. 415--426.

\bibitem[Liem and Panichella(2020)]{liem2020}
C.~C.~S. Liem and A.~Panichella, ``Run, forest, run? {O}n randomization and
  reproducibility in predictive software engineering,'' 2020, arXiv:2012.08387.

\bibitem[Herbold et~al.(2020)Herbold, Trautsch, and Trautsch]{herbold2020}
S.~Herbold, A.~Trautsch, and F.~Trautsch, ``On the feasibility of automated
  prediction of bug and non-bug issues,'' \emph{Empirical Software
  Engineering}, vol.~25, no.~6, pp. 5333--5369, 2020.

\bibitem[Jones et~al.(2015)Jones, Bradley, and Sakimura]{rfc7519}
M.~Jones, J.~Bradley, and N.~Sakimura, ``{JSON} {W}eb {T}oken ({JWT}),'' IETF,
  Tech. Rep. RFC 7519, 2015.

\bibitem[Hardt(2012)]{rfc6749}
D.~Hardt, ``The {OAuth} 2.0 authorization framework,'' IETF, Tech. Rep. RFC
  6749, 2012.

\bibitem[Krawczyk et~al.(1997)Krawczyk, Bellare, and Canetti]{rfc2104}
H.~Krawczyk, M.~Bellare, and R.~Canetti, ``{HMAC}: Keyed-hashing for message
  authentication,'' IETF, Tech. Rep. RFC 2104, 1997.

\bibitem[McCabe(1976)]{mccabe1976}
T.~J. McCabe, ``A complexity measure,'' \emph{IEEE Trans. Softw. Eng.}, vol.
  SE-2, no.~4, pp. 308--320, 1976.

\bibitem[Zhu et~al.(1997)Zhu, Hall, and May]{zhu1997coverage}
H.~Zhu, P.~A.~V. Hall, and J.~H.~R. May, ``Software unit test coverage and
  adequacy,'' \emph{ACM Comput. Surv.}, vol.~29, no.~4, pp. 366--427, 1997.

\bibitem[Nygard(2018)]{nygard2018release}
M.~T. Nygard, \emph{Release It!}, 2nd~ed.\hskip 1em plus 0.5em minus
  0.4em\relax Pragmatic Bookshelf, 2018.

\bibitem[Fowler(2014)]{fowler2014circuitbreaker}
M.~Fowler, ``{CircuitBreaker},'' 2014,
  \url{https://martinfowler.com/bliki/CircuitBreaker.html}.

\bibitem[Wessel et~al.(2018)Wessel, de~Souza, Steinmacher, Wiese, Polato,
  Chaves, and Gerosa]{wessel2018bots}
M.~Wessel, B.~M. de~Souza, I.~Steinmacher, I.~S. Wiese, I.~Polato, A.~P.
  Chaves, and M.~A. Gerosa, ``The power of bots: Characterizing and
  understanding bots in {OSS} projects,'' \emph{Proc. ACM Hum.-Comput.
  Interact.}, vol.~2, no. CSCW, pp. 182:1--182:19, 2018.

\bibitem[Vasilescu et~al.(2015)Vasilescu, Yu, Wang, Devanbu, and
  Filkov]{vasilescu2015ci}
B.~Vasilescu, Y.~Yu, H.~Wang, P.~Devanbu, and V.~Filkov, ``Quality and
  productivity outcomes relating to continuous integration in {GitHub},'' in
  \emph{Proc. ESEC/FSE}, 2015, pp. 805--816.

\bibitem[Mishra(2026)]{mishra2026zenodo}
S.~Mishra, ``Replication package: Boundary-mutation testing for pattern-based
  secret detection,'' 2026.

\end{thebibliography}
\end{document}